\documentclass[twocolumn]{aastex61}

\newcommand\aastex{AAS\TeX}

\received{July 1, 2016}
\revised{September 27, 2016}
\accepted{\today}
\submitjournal{ApJ}

\shorttitle{\aastex\ sample article}
\shortauthors{Hidalgo et al.}
\begin{document}

\title{The updated BaSTI stellar evolution models and isochrones: I. Solar scaled calculations}

\correspondingauthor{Sebastian L. Hidalgo}
\email{shidalgo@iac.es}

\author[0000-0002-0786-7307]{Sebastian L. Hidalgo}
\affil{Instituto de Astrof{\'i}sica de Canarias, Via Lactea s/n, La Laguna, Tenerife, Spain}
\affiliation{Department of Astrophysics, University of La Laguna, Via Lactea s/n, La Laguna, Tenerife, Spain}
%\nocollaboration

\author{Adriano Pietrinferni}
\affiliation{INAF - Osservatorio Astronomico d'Abruzzo, Via M. Maggini, s/n, I-64100, Teramo, Italy}
%\nocollaboration

\author{Santi Cassisi}
\affiliation{INAF - Osservatorio Astronomico d'Abruzzo, Via M. Maggini, s/n, I-64100, Teramo, Italy}
%\nocollaboration

\author{Maurizio Salaris}
\affiliation{Astrophysics Research Institute, 
Liverpool John Moores University, IC2, Liverpool Science Park, 146 Brownlow Hill, Liverpool, L3 5RF, UK}
%\nocollaboration

\author{Alessio Mucciarelli}
\affiliation{Dipartimento di Fisica e Astronomia - Universit{\'a} degli Studi di Bologna, Via Piero Gobetti 93/2, I-40129, Bologna, Italy}
\affiliation{INAF - Osservatorio di Astrofisica e Scienza dello Spazio di Bologna, via Piero Gobetti 93/3 - I-40129, Bologna, Italy}
%\nocollaboration

\author{Alessandro Savino}
\affiliation{Astrophysics Research Institute, Liverpool John Moores University, IC2, Liverpool Science Park, 146 Brownlow Hill, Liverpool, L3 5RF, UK}
\affiliation{Kapteyn Astronomical Institute, University of Groningen, Postbus 800, 9700 AV Groningen, The Netherlands}
%\nocollaboration

\author{Antonio Aparicio}
\affiliation{Instituto de Astrof{\'i}sica de Canarias, Via Lactea s/n, La Laguna, Tenerife, Spain}
\affiliation{Department of Astrophysics, University of La Laguna, Via Lactea s/n, La Laguna, Tenerife, Spain}
%\nocollaboration

\author{Victor Silva Aguirre}
\affiliation{Stellar Astrophysics Centre, Department of Physics and Astronomy, Aarhus University, Ny Munkegade 120, DK-8000 Aarhus C, Denmark}
%\nocollaboration

\author{Kuldeep Verma}
\affiliation{Stellar Astrophysics Centre, Department of Physics and Astronomy, Aarhus University, Ny Munkegade 120, DK-8000 Aarhus C, Denmark}
%\nocollaboration



\begin{abstract}
We present an updated release of the BaSTI ({\it a Bag of Stellar Tracks and Isochrones}) stellar model and isochrone library for 
a solar scaled heavy element distribution. The main input physics changed from the previous BaSTI release include the solar metal mixture, electron conduction opacities, 
a few nuclear reaction rates, bolometric corrections, and the treatment of the overshooting efficiency for shrinking convective cores.
The new model calculations cover a mass range between 0.1 and $15~M_{\odot}$, 22 initial chemical compositions 
between [Fe/H]=$-$3.20 and $+$0.45, with helium to metal enrichment ratio d$Y$/d$Z$=1.31. The isochrones cover an age range between 20~Myr and 14.5~Gyr,  
take consistently into account the pre-main sequence phase, and have been translated to a large number of popular photometric systems.
Asteroseismic properties of the theoretical models have also been calculated.
We compare our isochrones with results from independent databases and with several sets of observations, to test the accuracy of the calculations.
All stellar evolution tracks, asteroseismic properties and isochrones are made available through a dedicated Web site.
\end{abstract}


\section{Introduction} \label{sec:intro}

The interpretation of a vast array of astronomical observations, ranging from photometry and spectroscopy of galaxies and star clusters, to individual single and binary stars, 
to the detection of exoplanets, requires accurate sets of stellar model calculations covering all major evolutionary stages, and a 
wide range of mass and initial chemical composition.

Just as a few examples, the exploitation of the impressive amount of data provided by surveys like {\it Kepler} \citep[][-- asteroseismology]{kepler},  
APOGEE and SAGA \citep[][-- Galactic archaeology]{apogee, saga}, ELCID and ISLANDS 
\citep[][-- stellar population studies in resolved extra-galactic stellar systems]{elcid, islands}, present and future releases 
of the Gaia catalog \citep[see, e.g.,][]{gaia17}, observations with next-generation instruments like 
the James Webb Space Telescope and the Extremely Large Telescope, 
all require the use of extended grids of stellar evolution models.
In addition, the characterization of extrasolar planets in terms of 
their radii, masses, and ages \citep[the main science goal for example of the future PLATO mission, see][]{plato} 
is dependent on an accurate characterization 
of the host stars, that again requires the use of stellar evolution models. 

In the last decade several independent libraries of stellar models have been made available to the astronomical community, 
based on recent advances in stellar physics inputs like equation of state (EOS), Rosseland opacities, 
nuclear reaction rates. Examples of these libraries are 
BaSTI \citep{basti:04,basti:06,basti:09}, DSEP \citep{Dotter08}, Victoria-Regina \citep[see,][and references therein]{don:14}, Yale-Potsdam \citep{spada:17},
PARSEC \citep{parsec, chen14}, MIST \citep{mist}. 

Our group has built and delivered to the scientific community  
the BaSTI ({\it a Bag of Stellar Tracks and Isochrones}) stellar model and isochrone library, 
that has been extensively used to study field stars, stellar clusters, galaxies, both resolved and 
unresolved. In its first release, we delivered stellar models for a solar scaled heavy element mixture \citep{basti:04}, followed by 
complete sets of models for $\alpha-$enhanced \citep{basti:06} and CNO-enhanced heavy element
distributions \citep{basti:09}. In \cite{basti:13} we extended our calculations to the regime of extremely metal-poor and metal-rich chemical compositions. 
Extensions of the BaSTI evolutionary sequences to the final stages of the evolution of low- and intermediate-mass stars, i.e. the white dwarf cooling sequence and 
the asymptotic giant branch were published in \cite{bastiwd} and \cite{basti:07}, while sets of integrated properties and spectra self-consistently based on 
the BaSTI stellar model predictions were provided in \cite{percival:09}.

Since the first release of BaSTI, several improvements of the stellar physics inputs  
have become available, together with a number of revisions of the solar metal distribution, and corresponding revisions  
of the solar metallicity \citep[e.g.,][and references therein]{bergemann:14}. 
We have therefore set out to build a new release of the BaSTI library including these revisions of physics inputs and solar metal mixtures, 
still ensuring that our models satisfy a host of empirical constraints.
In addition --and this is entirely new compared to the previous BaSTI release-- we have also calculated and provide  
fundamental asteroseismic properties of the models.

This paper is the first one of a series that will present these new results. Here we focus on   
solar scaled non-rotating stellar models, while in a forthcoming paper we will publish $\alpha-$enhanced and $\alpha-$depleted 
models. Metal mixtures appropriate to study the multiple populations phenomenon in globular clusters 
\citep[see,][and references therein]{gratton:12, cs:13, piotto:15} will be presented in future publications. 

The plan of the paper is as follows. Section~\ref{code} details the 
physics inputs adopted in the new computations, including the new adopted solar heavy element 
distribution. 
Section~\ref{ssm} describes the standard solar model used to calibrate the mixing length and the He-enrichment ratio $\Delta{Y}/\Delta{Z}$, while Sect.~\ref{evo} 
presents the 
stellar model grid, the mass and chemical composition parameter space covered, the adopted bolometric corrections and the calculation of the 
asteroseismic properties of the models.
Section~\ref{databases} shows comparisons between our new models and recent independent calculations, 
whilst in Sect.~\ref{obs} the models are tested against a number of observational benchmarks.
Conclusions follow in Sect.~\ref{conclusions}.

\section{Stellar evolution code, solar metal distribution and physics inputs}
\label{code}

The evolutionary code\footnote{{\bf Starting from the work in preparation for the models published in   
\cite{basti:04}, we have adopted the acronym {\sl BaSTI} to identify both our own calculations and the stellar 
evolution code employed for these computations. The code is an independent evolution of the FRANEC code described in \cite{franec}. 
The current version is denoted as {\sl BaSTI version 2.0}.}} used in these calculations is the same one used to compute the original   
BaSTI library, albeit with several technical improvements to increase the model accuracy. 
For instance, we improved the mass layer (mesh) distribution and time step determinations, to obtain 
more accurate physical and chemical profiles for asteroseismic pulsational analyses.

The treatment of atomic diffusion of helium and metals has also been improved. We still include the effect of gravitational settling, chemical 
and temperature gradients (no radiative levitation) following \cite{thoul:94}, but the numerical treatment has been improved to ensure 
smooth and accurate chemical profiles for all the involved chemical species, from the stellar surface to the center.
{\bf We have also eliminated the traditional Runge-Kutta integration of the more external sub-atmospheric layers using the pressure 
as independent variable, with no energy generation equation and uniform chemical composition \citep[equal to the composition of the outermost 
layers integrated with the Henyey method, see e.g.][]{franec}. 
Historically this approach was chosen to save computing time, compared to a full Henyey integration up to the photosphere with mass as independent variable. 

This separate integration of the sub-atmosphere however prevents a fully 
consistent evaluation of the effect of atomic diffusion, that is included in the Henyey integration only. 
Depending on the selected total mass of the sub-atmospheric layers, the effect of diffusion on the surface abundances of low-mass stars can be appreciably underestimated.
In these new calculations we have included the sub-atmosphere in the Henyey integration, consisting typically of $\sim$300 mass layers. 
The more external mesh point contains typically a mass of the order of ${\rm 10^{-11}M_\odot}$. 

We have also performed tests to estimate the variation of the surface abundances of key elements when diffusion is treated with either pressure integration or Henyey 
mass integration of the sub-atmosphere. We fixed the total mass of the sub-atmospheric layers to $3.8 \times \ 10^{-5}$ times the total mass of the model, as in the previous BaSTI release.

In the case of a ${\rm 1M_\odot}$ model with solar initial metallicity and helium mass fraction  --$Z_{\odot}^{\rm ini}$=0.01721, $Y_{\odot}^{\rm ini}$=0.2695 (see Sect.~\ref{ssm})-- 
at the main sequence turn-off (approximately where the effect of diffusion is at its maximum) the surface mass fractions of He and Fe (representative of the metals) are essentially the same 
in both calculations. This is expected, given that the thickness of the sub-atmosphere is negligible compared to the total mass of the convective envelope.
Different is the case of lower metallicity low-mass models, with typically thinner (in mass) convective envelopes at the turn-off. A ${\rm 0.8M_\odot}$ model with 
initial $Z$=0.0001 and $Y$=0.247, displays at the turn-off an increase of the He and Fe mass fractions equal to 2\% and 4\% respectively, when the sub-atmosphere is included in the 
Henyey integration.}

\subsection{The solar heavy element distribution}
\label{distr}

The solar heavy element distribution sets the
zero point of the metallicity scale, and is also a critical input entering the calibration of the 
Solar Standard Model \citep[SSM--][]{vinyoles:17}, that in turn serves as calibrator of the mixing length parameter (see Sect.~\ref{mlt}), the initial solar 
He abundance and metallicity, and the d$Y$/d$Z$ He-enrichment ratio.

\lq{Classical}\rq\ estimates of the solar heavy element
distribution as that by \cite{gs:98} used in our previous BaSTI models, did allow SSMs to match very closely the 
constraints provided by helioseismology \citep[e.g.,][and references therein]{basti:04}. Recent reassessments by \cite{asp05} and 
\cite{asp09} have led to a downward revision of the solar metal abundances --by up to 40\% for
important elements such as oxygen. 
SSMs employing these new metal distributions produce a worse match to helioseismic constraints such as the sound speed  at the bottom of the convective envelope, as well as 
the location of the bottom boundary of surface convection, and the surface He abundance \citep[see, e.g.,][]{aldo09}.
This evidence has raised the so-called \lq{solar metallicity problem}\rq. A reanalysis of \cite{asp09} results and the use of 
an independent set of solar model atmospheres \citep[see, e.g.,][for a detailed discussion]{caffau:11} has provided a solar heavy element distribution 
intermediate between those by \cite{gs:98} and \cite{asp09}.

Although the problem is still unsettled and different solutions are under scrutiny \citep[see, e.g.,][]{vinyoles:17}, we decided to adopt the solar metal mixture by 
\cite{caffau:11}, supplemented when necessary by the abundances given by \cite{lodders:10}. The reference solar metal mixture 
adopted in our calculations is listed in Table~\ref{tab:mix}.
The actual solar metallicity is $Z_\odot = 0.0153$, while the corresponding actual ($Z/X$)$_\odot$ is equal to 0.0209.

\startlongtable
\begin{deluxetable}{c|cc}
\tablecaption{Abundances of the most relevant heavy elements in our adopted solar mixture\label{tab:mix}}
\tablehead{
\colhead{Element} & \colhead{Number fraction} & \colhead{Mass fraction} }
\startdata
C     &  0.260408    &   0.180125       \\
N     & 0.059656     &   0.048121      \\
O     & 0.473865     &   0.436614      \\ 
Ne   &  0.096751     &   0.112433     \\ 
Na   &   0.001681    &    0.002226    \\  
Mg   & 0.029899      &  0.041850      \\ 
Al     & 0.002487      &  0.003865      \\ 
Si     & 0.029218      &  0.047258     \\  
P      & 0.000237      &  0.000423     \\ 
S      &  0.011632     &   0.021476    \\  
Cl     & 0.000150      &  0.000306     \\ 
Ar     & 0.002727      &  0.006274     \\  
K      & 0.000106      &  0.000239    \\  
Ca    &  0.001760     &   0.004063    \\   
Ti      & 0.000072      &  0.000199     \\  
Cr     & 0.000385      &  0.001153    \\  
Mn    & 0.000266      &   0.000842   \\   
Fe     &  0.027268     &   0.087698   \\   
Ni      & 0.001431      &  0.004838    \\   
\enddata
\end{deluxetable}

\

\subsection{The treatment of convective mixing}

In our models --apart from the case of core He-burning in low- and intermediate-mass stars-- we use the Schwarzschild criterion to fix the formal 
convective boundary, plus instantaneous mixing in the convective regions. In case of models of massive stars, where 
layers left behind by shrinking convective cores during the 
main sequence (MS) have a hydrogen abundance that increases with increasing radius --formally requiring 
a semiconvective treatment of mixing-- we still use the Schwarzschild criterion  
and instantaneous mixing to determine the boundaries of the mixed region.   
This follows recent results from 3D hydrodynamics simulations of layered semiconvective regions \citep{w:13} that 
show how in stellar conditions, mixing in MS semiconvective regions is very fast and essentially equivalent 
to calculations employing the Schwarzschild criterion and instantaneous mixing \citep{moore:16}.

Theoretical simulations \citep[see, e.g.,][and references therein]{as:13, as:15, viallet:15},  
observations of open clusters and eclipsing binaries \citep[see, e.g.,][and references 
therein]{dsg94,magic:10,stancliffe:15, valle16, claret:16,claret}, as well as asteroseismic constraints \citep[see, e.g.,][]{victor:13}
show that in real stars chemical mixing beyond the formal convective boundary is required, and most likely results from the interplay of several physical processes,  
grouped in stellar evolution modelling under the generic terms {\sl overshooting} or {\sl convective boundary mixing}. 

In our calculations overshooting beyond the Schwarz\-schild boundary of MS convective cores is included as an 
instantaneous mixing between the formal convective border 
and layers at a distance $\lambda_{\rm ov} H_P$ from this boundary --keeping the radiative 
temperature gradient in this region. Here $H_P$ is the 
pressure scale height at the Schwarzschild boundary, and ${\rm \lambda_{OV}}$ a free parameter 
that we set equal to 0.2, decreasing to zero when the mass decreases below a certain value. This decrease is required because  
for increasingly small convective cores the Schwarzschild boundary moves progressively closer to the centre, and the 
local $H_P$ increases fast, formally diverging when the core shrinks to zero mass. Keeping ${\rm \lambda_{OV}}$  constant 
would produce increasingly large overshooting regions for shrinking convective cores.

How to decrease the overshooting efficiency is still somewhat arbitrary \citep[see, e.g.,][for a review of 
different choices found in the literature]{claret:16,screview:17}. As shown by \cite{basti:04}, the approach used 
to decrease the overshooting efficiency in the critical mass range between $\sim1.0~M_\odot$ 
and  $\sim1.5M_\odot$ has a potentially large effect on the isochrone morphology for ages around $\sim4-5$~Gyr \citep[see Fig.~1 in][]{basti:04}). 

\begin{figure}[ht!]
%\plotone{fig_overshoot.eps}
\begin{center}
 \includegraphics[width=3.2in]{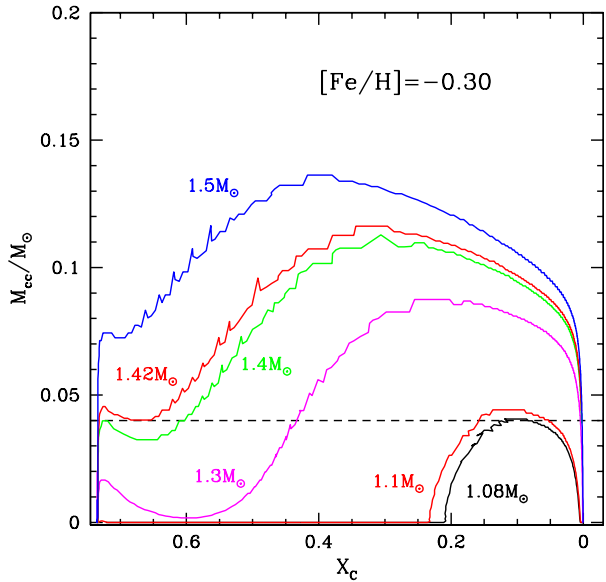} 
\caption{Convective core mass as a function of the central H mass fraction for stellar
models with the labelled masses,  and a metallicity Z=0.0077. The dashed line represents the value of 
$M_{cc}^{min}$ adopted in our calculations. 
In this example $M_{ov}^{inf}$=$1.08~M_\odot$ and $M_{ov}^{sup}$=$1.42~M_\odot$ (see text for details).
\label{fig:over}}
\end{center}
\end{figure}

In these new calculations we have chosen the following procedure to decrease  ${\rm \lambda_{OV}}$ with decreasing initial mass of the model.
For each chemical composition we have sampled the mass 
range between $1.0\le{M/M_\odot}\le1.5$ with a very fine mass spacing, and determined the initial mass 
($M_{ov}^{inf}$) that develops a convective core reaching at its maximum extension a mass  
$M_{cc}^{min}=0.04M_\odot$ during core H-burning. 
This initial mass is considered to be the maximum mass for models calculated  
with ${\rm \lambda_{OV}}$=0. We have then determined the minimum initial mass that develops 
a convective core always larger than $M_{cc}^{min}$ during the whole MS. 
This value of the initial mass is denoted as $M_{ov}^{sup}$. For models with initial masses equal or larger than $M_{ov}^{sup}$ 
we use ${\rm \lambda_{OV}}$=0.2, whereas between $M_{ov}^{inf}$ and $M_{ov}^{sup}$ the free parameter ${\rm \lambda_{OV}}$ increases linearly 
from 0 to 0.2.
An example of how we fix the values of $M_{ov}^{inf}$ and $M_{ov}^{sup}$ is shown in Fig.~\ref{fig:over}: For the selected metallicity
$M_{ov}^{inf}$ is equal to $1.08~M_\odot$, while $M_{ov}^{sup}$ is equal to $1.42~M_\odot$.

{\bf This criterion is obviously somehow arbitrary. It is based on numerical experiments we performed 
comparing the model predictions with empirical benchmarks such as eclipsing binaries and intermediate-age star clusters, as shown in Sect.~\ref{obs}.
Our choice indirectly introduces a dependence of $M_{ov}^{inf}$ and $M_{ov}^{sup}$ on the initial metallicity (see Table~\ref{tab:lib}).  
This is because the relationship between $M_{cc}^{min}$ and the total mass of the model depends on the efficiency of H-burning via the {\sl CNO-cycle}, that in turn is  
affected by a change of the absolute value of the total CNO abundance.}

The values of $M_{ov}^{inf}$ and $M_{ov}^{sup}$ for each initial chemical composition of our model grid are listed in Table~\ref{tab:lib}. 
This approach is different from the previous BaSTI release where, regardless of the chemical composition, 
we fixed the overshoot efficiency to its maximum value (${\rm \lambda_{OV}}$=0.2) 
for initial masses larger than or equal to $1.7~M_\odot$, decreasing linearly down to zero when the initial mass is equal to ${\rm 1.1M_\odot}$.

{\bf Before closing this discussion, it is interesting to compare our recipe for decreasing ${\rm \lambda_{OV}}$ with decreasing initial mass, 
with the results of a recent calibration by 
\cite{claret:16}. These authors compared their own model grid with effective temperatures and radii of a  
sample of detached double-lined eclipsing binaries with well determined masses, in the [Fe/H] range between about solar and $\sim -$1.01.  
They determined ${\rm \lambda_{OV}}$ equal to zero for masses lower than about $1.2M_\odot$, increasing 
to 0.2 in the mass range between $1.2M_\odot$ and $2M_\odot$. For masses larger than $2M_\odot$ ${\rm \lambda_{OV}}$ is equal to $\sim$0.2, as in our calculations. 
In the same metallicity range the value we adopt for $M_{ov}^{inf}$ ranges between $\sim$1.1 and $\sim1.2M_\odot$, whereas $M_{ov}^{sup}$ is always equal to $\sim$1.4$M_{\odot}$,  
about $0.6M_\odot$ smaller than \cite{claret:16} result. 
It is however very difficult to compare the two sets of results. Apart from possible intrinsic differences in the models, 
\cite{claret:16} determine from their fits also the individual values of the mixing length for each component,  
the initial metallicity $Z$ of each system, and allowed age differences up to 5\% between the components of each system.
They derived often systematically lower metallicites than the corresponding spectroscopic measurements.  
In Sect.~\ref{obs} we will see that 
our models fits well the mass-radius relationship of the systems KIC8410637 and OGLE-LMC-ECL-15260 \citep[this latter also studied by][]{claret:16} 
whose masses are in the 1.3-1.5~$M_{\odot}$ range, bracketing the upper limit where ${\rm \lambda_{OV}}$ reached 0.2 with our calibration.
We have imposed in our comparisons equal age for both systems, no variation of the mixing length and used models with chemical composition consistent with the spectroscopic measurements.}

In case of core He-burning of low- and intermediate-mass stars, we model core mixing with the semiconvective formalism 
by \cite{cctp:85}, and breathing pulses inhibited following \cite{caputo:89}.
During core He-burning in massive stars, we use the Schwarzschild criterion without overshooting to fix the boundary of the mixed region.

We do not include overshooting from the lower boundaries of convective envelopes.

\subsection{Radiative and electron conduction opacities}

The sources for the radiative Rosseland opacity are the same as for the previous BaSTI calculations. More in detail, 
opacities are from the OPAL calculations \citep{ir:96} for temperatures larger than ${\rm \log(T)=4.0}$, whereas calculations by \cite{ferguson} -- including 
contributions from molecules and grains -- have been adopted for lower temperatures. Both high- and low-temperature opacity tables 
have been computed for the solar scaled heavy element distribution listed in Table~\ref{tab:mix}. 

As for the electron conduction opacities, at variance with the models presented in \cite{basti:04,basti:06}, we have now adopted the results by \cite{cassisi:07}. 
As shown by \cite{cassisi:07}, these opacity calculations affect only slightly (small decrease) the He-core mass at
He-ignition for low-mass models, and the 
luminosity of the folllwing horizontal branch (HB) phase (small decrease), compared to 
the BaSTI calculations that were based on the \cite{pote:99} conductive opacities. For more details on this issue we refer the reader to the quoted reference as well as to 
\cite{serenelli:17}.

\subsection{Equation of state}

As in \cite{basti:04} we use the detailed EOS by A. Irwin\footnote{The EOS code is made publicly available at ftp://astroftp.phys.uvic.ca under the GNU General 
Public License.}. A brief discussion of the characteristics of this EOS can be found in \cite{csi:03}. We recomputed all required EOS tables 
for the heavy element distribution in Table~\ref{tab:mix} adopting the option \lq{\sl EOS1}\rq\ in Irwin's code. This option --recommended by A. Irwin 
\citep[see also the discussion in][]{csi:03}-- provides the best match to the OPAL EOS 
\citep{rogers:02}, and \cite{scvh} EOS in the low-temperature and high-density regime.

\subsection{Nuclear reaction rates}

The nuclear reaction rates are from the NACRE compilation \citep{nacre}, with the exception of the three following reactions, 
whose rates come from recent reevaluations:

\begin{itemize}

\item{${\rm ^3He(^4He,\gamma)^7Be}$ - \cite{cyburt:08};}

\item{${\rm ^{14}N(p,\gamma)^{15}O}$ -  \cite{formicola};}

\item{${\rm ^{12}C(\alpha,\gamma)^{16}O}$ - \cite{hammer}.}

\end{itemize}

The previous BaSTI calculations employed the NACRE rates \citep{nacre} for all reactions with the exceptions of the ${\rm ^{12}C(\alpha,\gamma)^{16}O}$
rate taken from \cite{kunz}

The first two reaction rates are important for H-burning; indeed the ${\rm ^{14}N(p,\gamma)^{15}O}$ reaction 
is crucial among those involved in the CNO-cycle, because it is the slowest one. 
The impact of this recent ${\rm ^{14}N(p,\gamma)^{15}O}$ rate on stellar evolution models has been investigated by \cite{imbriani},
\cite{weiss:05} and \cite{pietrinferni:10}. However, we have repeated here the analysis to verify the expected variation with respect to the 
previous BaSTI calculations, due to the combined effects of using the new rates for both ${\rm ^3He(^4He,\gamma)^7Be}$ and ${\rm ^{14}N(p,\gamma)^{15}O}$ nuclear reactions.
When all other physics inputs are kept fixed, we have found that:

\begin{itemize}

\item{for a ${\rm 0.8M_\odot}$, Z=0.0003 model, the luminosity at the MS turn-off (TO) increases by $\Delta\log(L/L_\odot)\sim 0.02$, 
while the age increases by about 210~Myr when passing from the NACRE  
reaction rates used in the previous BaSTI calculations to the ones adopted for the new models. For the same mass but a metallicity Z=0.008 
the effects are smaller, with a MS TO luminosity increased by about 0.01~dex and an age increased by $\sim 30$~Myr;}

\item{as for the evolution along the red giant branch (RGB), the effect of the new rates on the RGB bump luminosity is 
completely negligible at Z=0.008, while the RGB bump luminosity increases by $\Delta\log(L/L_\odot)\sim 0.04$ at Z=0.0003.
Regardless of the metallicity, the use of the new rates decreases the RGB tip brightness by $\Delta\log(L/L_\odot)\sim 0.02$ 
in agreement with the results by \cite{pietrinferni:10} and \cite{serenelli:17}.}

\end{itemize}

The ${\rm ^{12}C(\alpha,\gamma)^{16}O}$ reaction is one of the most critical nuclear processes in stellar astrophysics, because of its impact on a 
number of astrophysical 
problems \citep[see, e.g.,][and references therein]{csi:03, cs:13}. The more recent assessment of this reaction rate is
not significantly different 
from \cite{kunz} as used by \cite{basti:04}. As a consequence, the use of this new rate has a small impact on the models: For instance,  
the core He-burning lifetime is decreased by a negligible $\sim 0.2$\% when using this new rate compared to models calculated with the older \cite{kunz} rate.

As in the previous BaSTI calculations, 
electron screening is calculated according to the appropriate choice between strong, intermediate, and weak, following \cite{scree1} and \cite{scree2}.

\subsection{Neutrino energy losses}

Neutrino energy losses are included with the same prescriptions as in the previous BaSTI calculations. 
For plasma neutrinos we use the rates by \cite{hrw94}, supplemented by \cite{munak} rates for the other relevant 
neutrino production processes.

\subsection{Superadiabatic convection and outer boundary conditions}
\label{mlt}

The combined effect of the treatment of the superadiabatic layers 
of convective envelopes, and the method to determine the outer boundary conditions of the models, has a major 
impact on the effective temperature scale of stellar models with deep convective envelopes (or fully convective).

As in the previous BaSTI models, we treat the superadiabatic convective layers according to the \cite{bv:58} flavor of the 
mixing length theory, using the formalism by \cite{cg:68}. The value of the mixing length parameter ${\rm \alpha_{ML}}$ is fixed by 
the solar model calibration {\bf to 2.006 (see Sect.~\ref{ssm} for more details) and kept the same for all masses, initial chemical compositions and evolutionary phases}.

Regarding the outer boundary conditions, in the previous BaSTI models they were obtained by integrating the atmospheric layers employing 
the $T(\tau)$ relation provided by \cite{ks:66}. In this new release we decided to employ the alternative solar semi-empirical 
$T(\tau)$ by \cite{vernazza:81}. More specifically, we implemented in our evolutionary code the following fit to the tabulation provided by \cite{vernazza:81}:

\begin{equation}
T^4 =0.75 \ T_{eff}^4 \ (\tau+1.017-0.3 e^{-2.54 \tau}-0.291 e^{-30 \tau})
\label{eq:ttau}
\end{equation}

\begin{figure}[ht!]
%\plotone{fig_085ttau.eps}
\begin{center}
\includegraphics[width=3.2in]{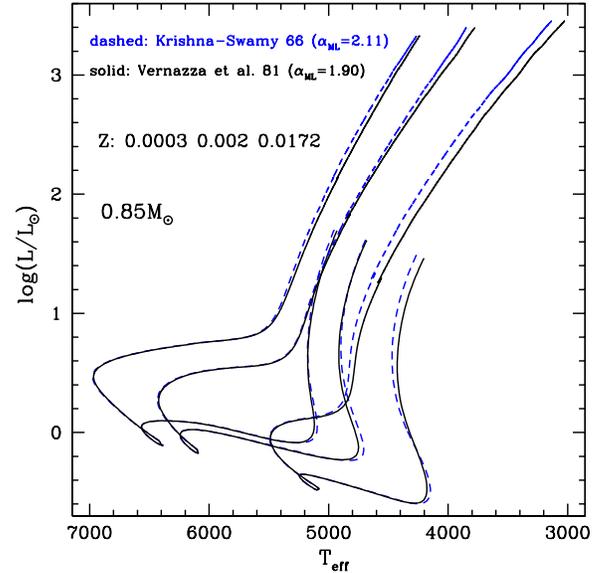} 
\caption{HRDs of models computed with 
two different assumptions about the $T(\tau)$ 
relation used to calculate the outer boundary conditions, for the labelled mass and metallicities. 
{\bf The solar calibrated mixing length values for each choice of the  $T(\tau)$ 
relation are also shown}.\label{fig:ttau}}
\end{center}
\end{figure}

As shown by \cite{sc:15}, model tracks computed with this $T(\tau)$ relation approximate well results obtained using the 
hydro-calibrated $T(\tau)$ relationships determined from the 3D radiation hydrodynamics calculations by \citet{trampedach14} for the solar chemical composition. 
Figure~\ref{fig:ttau} shows the Hertzsprung-Russell diagram (HRD) of 
$0.85~M_\odot$ evolutionary tracks from the pre-MS to the tip of the RGB,  
computed for three labelled initial metallicities. The physics inputs are kept the same as the old BaSTI calculations,  
but for the $T(\tau)$ relation, that is either from \cite{ks:66} or \cite{vernazza:81}.  
For both choices the value of ${\rm \alpha_{ML}}$ has been fixed by an appropriate solar calibration.

The two sets of models overlap almost perfectly along the MS at all $Z$, 
whereas some differences in $T_{eff}$ at fixed luminosity appear along the RGB (and the pre-MS).  
Differences are of about 60~K at the lowest metallicity, reaching $\sim90$~K at solar metallicity. 
Tracks calculated with the \cite{vernazza:81} $T(\tau)$ are always the cooler ones. 
For a more detailed discussion on the impact of different $T(\tau)$ relations on 
the $T_{eff}$ scale of RGB stellar models we refer to \cite{sc:15} and references therein.

In the first release of BaSTI the minimum stellar mass was set to $0.50~M_\odot$ for all chemical compositions, while 
these new calculations include the mass range below $0.50~M_\odot$, down to $0.10~M_\odot$. 
As extensively discussed in the literature \citep[see, e.g.,][and references therein]{baraffe:95,  allard:97, brocato:98, chabrier:00} 
in this regime of so-called very low-mass (VLM) stars, {\bf i.e. ${\rm M\le0.45M_\odot}$}, outer boundary conditions provided by accurate non-gray model atmospheres are required. 
Therefore for the VLM model 
calculations we employed boundary conditions (pressure and temperature at a Rosseland optical depth $\tau$=100) taken from 
the {\it PHOENIX} model atmosphere library\footnote{The model atmosphere dataset is publicly available at the following URL: 
http://phoenix.ens-lyon.fr/Grids/ } \citep[][and references therein]{allard:12}, more precisely the {\it BT-Settl} model set.
These model atmospheres properly cover the required parameter space in terms of effective temperature, surface gravity and metallicity range. However, this set of 
models have been computed for the \cite{asp09} solar heavy element distribution, that is different from the one adopted in our calculations (see Sect.~\ref{distr}). 

\begin{figure}[ht!]
%\plotone{fig_vlmmix.eps}
\begin{center}
\includegraphics[width=3.4in]{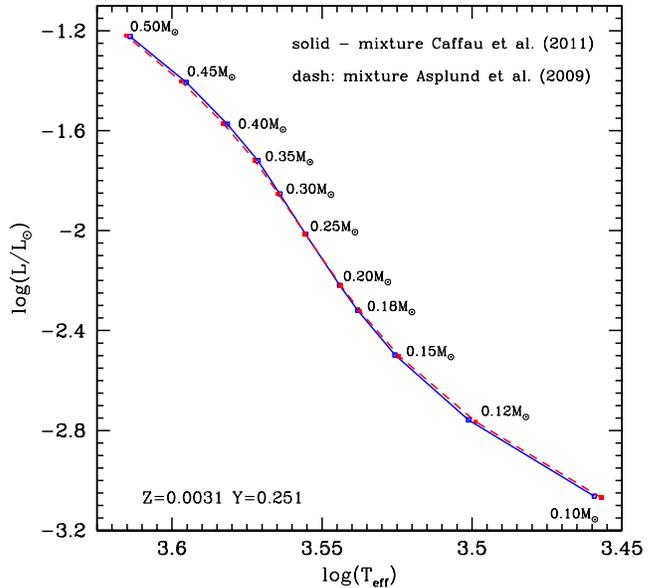}
\caption{HRD of core H-burning models for an age of 10~Gyr and the labelled initial chemical composition {\bf and masses}. 
Boundary conditions have been obtained from model atmospheres calculated using the 
labelled solar heavy element mixtures (see text for details).\label{fig:vlmmix}}
\end{center}
\end{figure}

One could argue that this difference in the heavy element mixture may have an impact on the predicted spectral energy distribution, but it should have only a minor 
effect on the model atmosphere structure, hence on the derived outer boundary conditions. We have verified this latter point as follows. 
The {\it PHOENIX} model atmosphere repository contains a subset of models --labelled CIFIST2011 -- computed with the same solar heavy element 
distribution as in our calculations \citep{caffau:11}, for a few selected metallicities. 
We have calculated sets of VLM models using alternatively the PHOENIX boundary conditions 
for the \cite{asp09} mixture and the \cite{caffau:11} one. Figure~\ref{fig:vlmmix} shows the result of such comparison for one selected metallicity. As expected the 
the two sets of VLM calculations provide very similar HRDs. Differences in bolometric luminosity and effective temperature are
vanishing small for masses larger than $\sim 0.12~M_\odot$, while they are equal to 
just $\Delta\log(L/L_\odot)\sim 0.007$ and $\Delta{T_{eff}}\sim 16$~K, for smaller masses.

{\bf We close this section with more details about the transition from VLM models with outer boundary conditions determined from {\it PHOENIX} model atmospheres, to    
models calculated with the $T(\tau)$ relation in Eq.~\ref{eq:ttau}. 
To achieve a smooth transition in the $log(L/L_\odot) - T_{eff}$ diagram between the two regimes, for each chemical composition we computed models   
with mass up to $0.70~M_\odot$ with the {\it PHOENIX} boundary conditions, and models with mass down to $0.4~M_\odot$ using the $T(\tau)$ relation.
In the overlapping mass range we selected a specific transition mass corresponding to the pair of models 
--that happen to fall in the range between $\sim~0.5M_\odot$ and $\sim 0.65~M_\odot$, 
depending on the initial composition-- showing negligible differences in both bolometric luminosity and effective temperature, 
typically $\Delta{T_{eff}}\le25$~K, and $\Delta\log(L\L_\odot)\le0.004$. For masses equal and lower than this mass 
we keep the calculations with {\it PHOENIX} boundary conditions, and above this limit the models with $T(\tau)$ integration. 
This allows to calculate isochrones displaying a smooth transition between the two boundary condition regimes.}

\subsection{Mass loss}

Mass loss is included with the \cite{reimers} formula, as in the previous BaSTI models. The free parameter $\eta$ entering this mass loss prescription 
has been set equal to 0.3, following the {\sl Kepler}  observational constraints discussed in \cite{miglio:12}. 
We provide also stellar models computed without mass loss ($\eta$=0).
The previous BaSTI calculations included three options, $\eta$=0, 0.2 and 0.4, respectively\footnote{The release of the previous BaSTI models with $\eta=0$ is 
not directly available at the old URL site, but can be obtained on demand.}.

\section{The Standard Solar Model}
\label{ssm}

As already mentioned, the calibration of the SSM sets the value of ${\rm \alpha_{ML}}$, and the initial solar He abundance and metallicity. 
At the solar age \citep[$t_\odot=4.57$~Gyr][]{bahcall:95} 
our $1~M_\odot$ SSM (including diffusion of both He and metals and calculated starting from the pre-main sequence)
matches luminosity, 
radius \citep[$L_\odot=3.842\times10^{33}$~erg/s and $R_\odot=6.9599\times10^{10}$~cm, respectively, as given by 
][]{bahcall:95}, and the present ($Z/X$)$_\odot$ \citep{caffau:11} abundance ratio with  
initial abundances $Z^{\rm ini}_\odot=0.01721$ and $Y^{\rm ini}_\odot=0.2695$, and mixing length ${\rm \alpha_{ML}=2.006}$.

Our SSM has a surface He abundance $Y_{\rm \odot, surf}=$0.238 and a radius of the boundary of the surface convective zone $R_{\rm CZ}/R_\odot$ equal to 0.722. 
These values have to be compared with the asteroseismic estimates 
$R_{\rm CZ}/R_\odot=0.713\pm0.001$ \citep{basu:97} and $Y_{\rm \odot, surf}=0.2485\pm0.0035$ \citep{basu:04}.
These differences between models and observations are common to all SSMs based on the revised solar surface compositions discussed before 
\citep[e.g.][and references therein]{basu:04, vinyoles:17}.
Differences are larger when using the lower $Z$ solar abundances by \cite{asp09}, as discussed by \cite{mist}. 
This is an open problem, and efforts are being devoted 
to explore the possibility of suitable changes to the SSM input physics, such as radiative opacities 
\citep[we refer to][for a detailed analysis of this issue]
{villante:10, krief:16, vinyoles:17}.

\section{The stellar model library}
\label{evo}

Our new model library increases significantly the number of available metallicities, compared to the old BaSTI calculations.
We have calculated models for 22 metallicities ranging from $Z=10^{-5}$ up to $\sim 0.04$; the exact values are listed 
in Table~\ref{tab:lib}. We adopted a primordial He abundance $Y=0.247$ based on the cosmological baryon density following {\sl Planck} 
results \citep{coc}. With this choice of the primordial He abundance and the initial solar He abundance obtained from the SSM calibration 
we obtain an He-enrichment ratio d$Y$/d$Z$=1.31, that we have used in our model grid computation. 
For each metallicity, the corresponding initial He abundance and [Fe/H] are listed in Table~\ref{tab:lib}.

\subsection{Evolutionary tracks}

{\bf As with the first release of the BaSTI database, we have calculated several model grids by varying once at a time some modelling assumptions. 
A schematic overview of all grids made available in the new BaSTI repository is provided in Table~\ref{tab:grids}.
Our reference set of models is set a) in Table~\ref{tab:grids}, that include main sequence convective core overshooting, 
mass loss with $\eta$=0.3 and atomic diffusion of He and metals.} 

For each chemical composition (and choice of modelling assumptions) we have computed 56 evolutionary sequences. 
The minimum initial mass is $0.1~M_\odot$, while the maximum value is  
$15~M_\odot$). For initial masses below $0.2~M_\odot$ we computed evolutionary tracks for masses equal to 0.10, 0.12, 0.15 and 
$0.18~M_\odot$. In the range between 0.2 and $0.7~M_\odot$ a mass step equal to $0.05~M_\odot$ has been adopted.
Mass steps equal to $0.1~M_\odot$, 
$0.2~M_\odot$, 0.5~$M_\odot$ and 1~$M_\odot$ have been adopted for the mass ranges $0.7-2.6~M_\odot$, $2.6-3.0~M_\odot$, 
$3.0-10.0~M_\odot$, and masses larger than $10.0M_\odot$, respectively.

Models less massive than $4.0~M_\odot$ have been computed from the pre-MS, whereas more massive models have been computed starting from a chemically 
homogeneous configuration on the MS. Relevant to pre-MS calculations, 
the adopted mass fractions for D, ${\rm ^{3}He}$ and ${\rm ^{7}Li}$  are equal to $3.9 \ 10^{-5}$, $2.3 \ 10^{-5}$, and $2.6 \ 10^{-9}$ respectively.

All stellar models -- but the less massive ones whose core H-burning lifetime is longer than the 
Hubble time -- 
have been calculated until the start of the thermal pulses (TPs)\footnote{In the near future we plan to extend these computations to the end of the  
TP phase using the 
synthetic AGB technique \citep[see, e.g.,][and references therein]{basti:07}.} on the Asymptotic Giant Branch (AGB), or C-ignition for the more massive ones.
For the long-lived low-mass models we have stopped the calculations when the central H mass fraction is $\sim$0.3 (corresponding to ages already much 
larger than the Hubble time).

For each initial chemical composition we provide also an extended set of core He-burning models suited to study 
the HB in old stellar populations. 
We have considered various values of the total mass \citep[with a fine mass spacing, as in][]{basti:04}
but the same mass for the He-core and the same envelope chemical stratification, corresponding to a RGB progenitor at the He-flash 
for an age of $\sim12.5$~Gyr. 

\begin{figure}[ht!]
%\plotone{fig_kpt.eps}
\begin{center}
\includegraphics[width=3.2in]{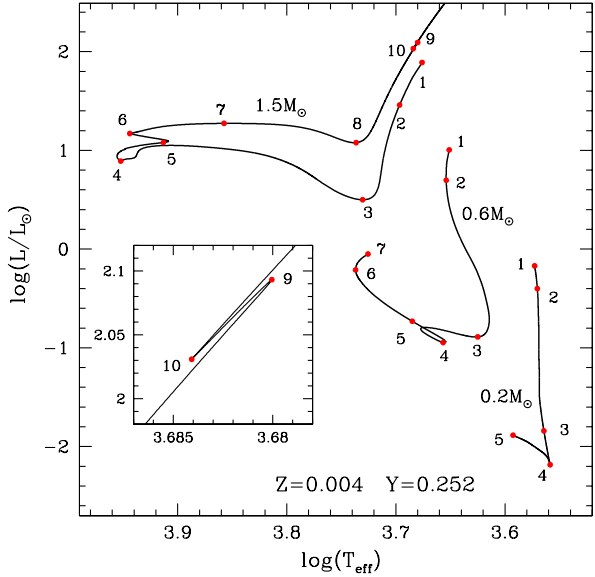}
\caption{HRD of selected evolutionary tracks and the labelled initial chemical composition.  
We also show the position of the first 10 key points used to normalize the tracks. 
The inset is an enlargement of the RGB bump phase, to show the exact position of key points 9 and 10.\label{fig:key}}
\end{center}
\end{figure}

All evolutionary tracks presented in this work have been reduced to the same number of points (\lq{normalized}\rq ) to calculate isochrones 
\citep[see, e.g.,][for a discussion on this issue]{dott} and for ease of 
interpolation, by 
adopting the same approach extensively discussed in \cite{basti:04} and updated in 
\cite{basti:06}. This method is based on the identification of some characteristic homologous points (keypoints) 
corresponding to well-defined evolutionary stages along each individual track  
\citep[see][for more details on this issues]{basti:04}. Given that almost all the evolutionary tracks now include the pre-MS stage, we 
added three additional keypoints compared to the previous BaSTI calculations. 
The first one is taken at an age of $10^4$~yr, the second one corresponds to the end of the deuterium burning stage, while
the third keypoint is set at the first minimum of the surface luminosity for all models but the VLM ones. For these latter masses this point corresponds to the stage when
the energy produced by the {\sl p-p chain} starts to dominate the energy budget. The fourth keypoint corresponds to the zero age main sequence (ZAMS) 
{\bf defined as the model fully sustained by nuclear reactions, with all secondary elements at their equilibrium abundances\footnote{This 
stage also corresponds to the minimum luminosity during the core H-burning stage.}.} 
However, for VLM models that attain nuclear equilibrium of the secondary elements involved in the {\sl p-p chain} over extremely long timescales,  
this keypoint corresponds to the first minimum of the bolometric luminosity.
All subsequent keypoints are fixed exactly as in the previous BaSTI database. Table~\ref{tab:key} lists the correspondence between 
keypoints and evolutionary stages as well as the corresponding line number in the normalized evolutionary track, 
%\footnote{As discussed we computed the Pre-MS stage 
%for all stellar structures with mass ${\rm \le 4.0M_\odot}$; therefore in order to allow an easy interpolation on the whole set of evolutionary tracks, for stellar masses 
%larger than the quoted limit we added,  before the line corresponding to the ZAMS stage, 99 fictitious lines -- obtained from the pre-ZAMS relaxation phase.}; 
while 
Fig.~\ref{fig:key} shows the location of a subset of  keypoints (the first ten ones) on selected evolutionary tracks.

For each chemical compositions, these normalized evolutionary tracks are used to compute 
extended sets of isochrones for ages between 20~Myr and 14.5~Gyr (older isochrones can be also computed on demand). 

\startlongtable
\begin{deluxetable}{ccccc}
\tablecaption{Grid of initial chemical abundances and corresponding values (in solar masses) of $M_{ov}^{inf}$ and $M_{ov}^{sup}$  
(see text for details).\label{tab:lib}}
\tablehead{
\colhead{$Z$} & \colhead{$Y$} & \colhead{[Fe/H]} & \colhead{$M_{ov}^{inf}$}  & \colhead{$M_{ov}^{sup}$} }
\startdata
   0.00001 &  0.2470 &  $-3.20$  & 1.30  & 2.09\\   
   0.00005 &  0.2471 &  $-2.50$  & 1.30  & 1.78\\    
   0.00010 &  0.2471 &  $-2.20$  & 1.30  & 1.68\\    
   0.00020 &  0.2472 &  $-1.90$  & 1.30  & 1.59\\    
   0.00031 &  0.2474 &  $-1.70$  & 1.30  & 1.54\\    
   0.00044 &  0.2476 &  $-1.55$  & 1.30  & 1.50\\    
   0.00062 &  0.2478 &  $-1.40$  & 1.32  & 1.47\\    
   0.00079 &  0.2480 &  $-1.30$  & 1.32  & 1.45\\    
   0.00099 &  0.2483 &  $-1.20$  & 1.24  & 1.44\\    
   0.00140 &  0.2488 &  $-1.05$  & 1.21  & 1.43\\    
   0.00197 &  0.2496 &  $-0.90$  & 1.17  & 1.42\\    
   0.00311 &  0.2511 &  $-0.70$  & 1.13  & 1.42\\    
   0.00390 &  0.2521 &  $-0.60$  & 1.10  & 1.42\\    
   0.00614 &  0.2550 &  $-0.40$  & 1.09  & 1.42\\    
   0.00770 &  0.2571 &  $-0.30$  & 1.08  & 1.42\\    
   0.00964 &  0.2596 &  $-0.20$  & 1.08  & 1.42\\   
   0.01258 & 0.2635  &  $-0.08$  & 1.08  & 1.43\\
   0.01721 &  0.2695 &  $ 0.06$  & 1.09  & 1.43\\	   
   0.02081 &  0.2742 &  $ 0.15$  & 1.11  & 1.47\\	   
   0.02865 &  0.2844 &  $ 0.30$  & 1.10  & 1.42\\	   
   0.03905 &  0.2980 &  $ 0.45$  & 1.09  & 1.40\\	   
\enddata
\end{deluxetable}

\begin{deluxetable}{cccc}
\tablecaption{The various grids of stellar models provided in the database. \label{tab:grids}}
\tablehead{
\colhead{Case} & \colhead{Convective overshooting} & \colhead{Mass loss efficiency}  & \colhead{Diffusion} }
\startdata
   a &  Yes &  $\eta=0.3$  & Yes\\   
   b &  Yes &  $\eta=0.3$  & No\\   
   c &  Yes &  $\eta=0.0$  & No\\   
   d &  No &  $\eta=0.0$  & No\\   
\enddata
\end{deluxetable}

\begin{deluxetable*}{crc}
\tablecaption{Correspondence between evolutionary stage, key point and line number of the normalized tracks.}

\label{tab:key}
\tablehead{
\colhead{Key Point} &  \colhead{Line} & \colhead{Evolutionary Phase}   }
\startdata
   1 &  1 & Age equal to 1000~yr \\   
   2 &  20 & End of deuterium burning \\    
   3 &  60 & The first minimum in the surface luminosity, or when nuclear energy starts to dominate the energy budget \\    
   4 &  100 & Zero age main sequence or minimum in bolometric luminosity for VLM models \\    
   5 &   300 & First minimum of $T_{eff}$ for high-mass or central H mass fraction $X_{c}$=0.30 for low-mass and VLM models\\    
   6 &   360 & Maximum in $T_{eff}$ along the MS (TO point)\\    
   7 &   420 &Maximum in $\log(L/L_\odot)$ for high-mass or $X_{c}$=0.0 for low-mass models\\    
   8 &   490 &Minimum in $\log(L/L_\odot)$ for high-mass or base of the red giant branch for low-mass models\\    
   9 &   860 &Maximum luminosity along the RGB bump\\    
  10 &   890 &Minimum luminosity along the RGB bump\\    
  11 &   1290 &Tip of the RGB\\    
  12 &   1300 & Start of quiescent core He-burning\\    
  13 &   1450 &Central abundance of He equal to 0.55\\    
  14 &   1550 & Central abundance of He equal to 0.50\\    
  15 &   1650 &Central abundance of He equal to 0.40\\    
  16 &   1730 &Central abundance of He equal to 0.20\\   
  17 &   1810 & Central abundance of He equal to 0.30\\       
  18 &   1950 & Central abundance of He equal to 0.00\\       
  19 &   2100 & The energy associated to the {\sl CNO-cycle} becomes larger than that provided by He-burning\\       
\enddata
\end{deluxetable*}

\begin{figure*}[ht!]
%\plotone{fig_kpt.eps}
\begin{center}
\includegraphics[width=7.0in]{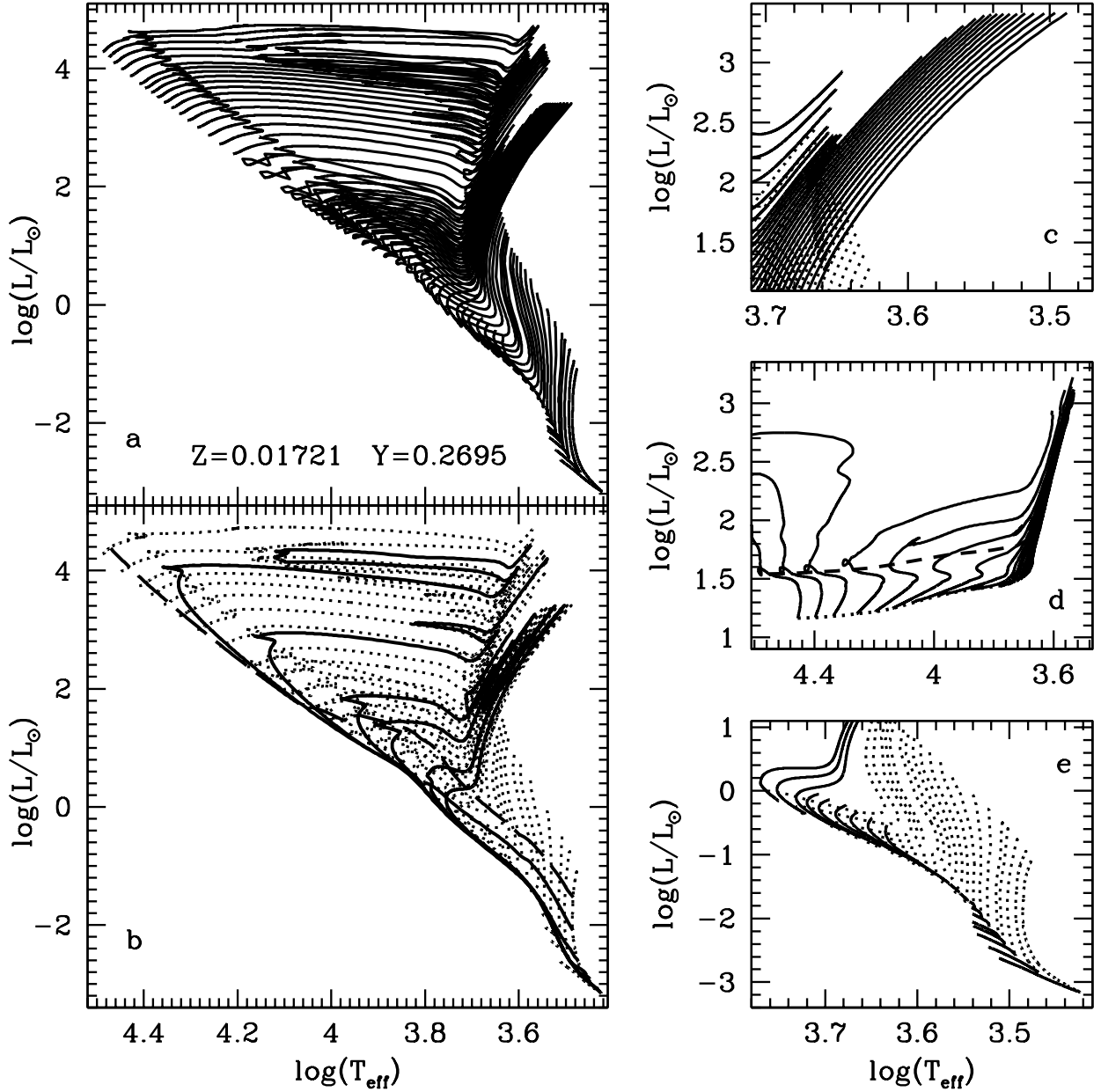}
\caption{HRDs of the full set of reference tracks and isochrones calculated for the labelled initial chemical composition (panel $a$), 
and a subset of isochrones {\bf for 5~Myr (long dashed line), and} 20~Myr, 100~Myr, 500~Myr, 1~Gyr, 4~Gyr and 14~Gyr, solid lines in panel $b$) overlaid onto the track grid (dashed lines). 
Panel $c$ shows selected RGB tracks (solid lines) and part of their pre-MS evolution (dotted lines),
while panel $d$ displays the full set of HB tracks. The zero age HB 
is shown as a dotted line, while the dashed line corresponds to central He exhaustion. Panel $e$ displays a subset of pre-MS
(dotted), MS and RGB tracks with mass 
between 0.1 a 1.0~$M_{\odot}$ (see text for details).\label{fig:database}}
\end{center}
\end{figure*}

Figure~\ref{fig:database} shows an example of the full set of reference tracks and isochrones calculated for one chemical composition 
(Y=0.2695, Z=0.01721). 
Panel $a$ displays the full grid of tracks for masses ranging from 0.1$M_{\odot}$ to 15$M_{\odot}$, while 
panel $c$ focuses on the RGB region for 
a subset of models with mass between 0.4 and 4.5$M_{\odot}$ (dotted lines denote the pre-MS evolution of the same models).
The set of HB tracks is shown in panel $d$, for a RGB progenitor mass equal to 1.0$M_{\odot}$,  
and minimum HB mass equal to 0.4727$M_{\odot}$, while panel $e$ displays a subset of pre-MS, MS and RGB tracks with mass 
between 0.1 a 1.0$M_{\odot}$. Finally, panel $b$ displays 
a set of isochrones with ages equal to 20~Myr, 100~Myr, 500~Myr, 1~Gyr, 4~Gyr and 14~Gyr, respectively (solid lines), 
overlaid onto the full set of tracks (dashed lines). 

\subsection{Bolometric corrections}

Bolometric luminosities and effective temperatures along evolutionary tracks and isochrones need to be translated to 
magnitudes and colors in sets of photometric filters, for comparisons with observed color-magnitude-diagrams (CMDs), and to predict integrated fluxes 
of unresolved stellar populations. This requires sets of stellar spectra covering 
the relevant parameter space in terms of metallicity, surface gravity and
effective temperature of the models. 
For such aim, a new grid of model atmospheres has been computed using the latest version 
of the ATLAS9 code\footnote{http://wwwuser.oats.inaf.it/castelli/sources/atlas9codes.html} 
originally developed by R. L. Kurucz \citep{kurucz70}.
ATLAS9 allows to calculate one-dimensional, plane-parallel model atmospheres 
under the assumption of local thermodynamical equilibrium for all the species. 
The method of the opacity distribution function \citep[ODF --][]{kurucz:74} is employed to handle the 
line opacity, by pretabulating the line opacity as a function of gas pressure and 
temperature in a given number of wavelength bins. ODFs and Rosseland mean opacity 
tables are calculated for a given metallicity (fixing the chemical mixture) and 
for a given value of microturbulent velocity. 
Even if the computation of ODFs can be time consuming, 
the calculation of any model atmosphere (defined by its effective temperature and gravity) for the metallicity 
and microturbulent velocity corresponding to the adopted ODF turns out to be very fast.

Grids of ATLAS9 model atmospheres based on suitable ODFs are freely available but 
based on different solar chemical abundances compared to the one used in our calculations. 
The grid by \citet{castelli04} 
adopted the solar abundances by \citet{gs:98}, that computed by \citet{kirby11} 
the abundances by \citet{ag89}, while the recent one by \citet{meszaros12} 
for the APOGEE survey used the abundances by \citet{asp05}. 
For the new grid presented here we adopted the same solar metal distribution of the stellar 
evolution calculations.
For the computation of new ODFs, Rosseland opacity tables and model atmospheres 
we followed the scheme described in \citet{meszaros12}.

For each [Fe/H] and microturbulent velocity, one ODF and one Rosseland 
opacity table are calculated using the codes DFSYNTHE and KAPPA9 \citep{castelli05}, 
respectively. The [Fe/H] grid ranges from $-$4.0 to +0.5~dex 
in steps of 0.5~dex from $-$4.0 to $-$3.0 dex, and in steps of 0.25~dex for the other 
metallicities, assuming solar scaled abundances for all elements. 
The adopted values for the microturbulent velocities are 0, 1, 2, 4 and 8~km/s. 
In the calculation of the ODFs we included all atomic and molecular transitions 
listed in F. Castelli website\footnote{http://wwwuser.oats.inaf.it/castelli/linelists.html}; 
in particular the linelist for TiO is from \citet{schwenke98} and that 
for ${\rm H_{2}O}$ is from \citet{langhoff}.

For each [Fe/H] (but adopting only the microturbulent velocity of 2~km/s) 
a grid of ATLAS9 model atmospheres has been computed, covering the  
effective temperature-surface gravity parameter space summarized in Table~\ref{tab:gridmod},  
for a total of 475 models.

Similarly to those computed by \citet{castelli04}, these new model atmospheres 
include 72 plane-parallel layers ranging from $\log{\tau}$=$-$6.875 (where $\tau$ is the Rosseland optical depth) to 
+2.00, in steps of 0.125, and have been computed with the overshooting option 
switched off, adopting a mixing-length equal to 1.25 as previous calculations.
For each model atmosphere, the corresponding emerging flux has been then computed.

\begin{deluxetable}{cccc}
\tablecolumns{4} 
\tablewidth{0pc}  
\tablecaption{Effective temperature and surface gravity ranges covered by our new grid of ATLAS9 model atmospheres and spectra, together 
with the grid spacings $\Delta_{T_{eff}}$ and $\Delta_{log(g)}$.\label{tab:gridmod}}
\tablehead{ 
\colhead{$T_{eff}$}  &  \colhead{$\Delta_{T_{eff}}$}  &     \colhead{log(g)}  &  \colhead{$\Delta_{\rm log(g)}$} \\
 (K)  &  (K)  &  (c.g.s)  &  (c.g.s)  \\}
\startdata 
\hline   
 3500--6000     &  250     &  0.0--5.0   &  0.5  \\  
 6250--7500     &  250     &  0.5--5.0   &  0.5  \\  
 7750--8250     &  250     &  1.0--5.0	 &  0.5  \\  
 8500--9000     &  250     &  1.5--5.0	 &  0.5  \\  
9250--11750     &  250     &  2.0--5.0	 &  0.5  \\  
12000--13000    &  250     &  2.5--5.0	 &  0.5  \\  
13000--19000    & 1000     &  2.5--5.0	 &  0.5  \\  
20000--26000    & 1000     &  3.0--5.0   &  0.5  \\  
27000--31000    & 1000     &  3.5--5.0   &  0.5  \\  
32000--39000    & 1000     &  4.0--5.0   &  0.5  \\  
40000--49000    & 1000     &  4.5--5.0   &  0.5  \\  
50000           &   ---    &   5.0       &  ---  \\ 
\hline
\enddata 
\end{deluxetable}

\begin{figure}[ht!]
%\plotone{grid.eps}
\begin{center}
\includegraphics[width=3.4in]{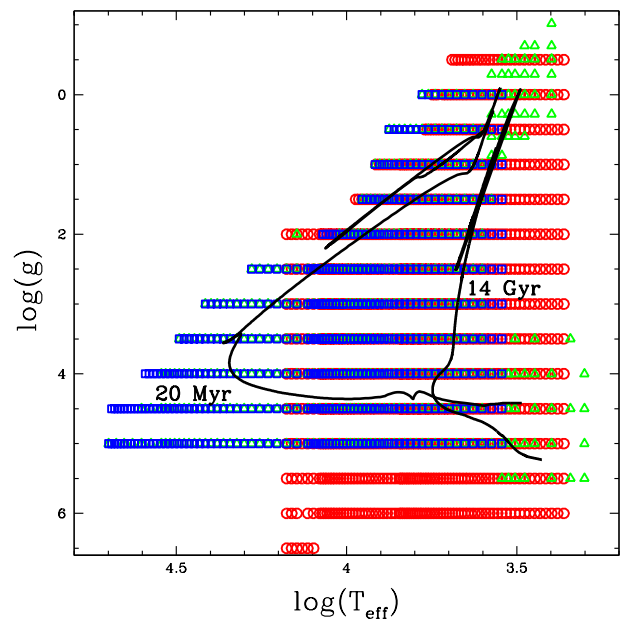}
\caption{The $T_{eff} -$ log(g) coverage ([Fe/H] = 0) of the adopted spectral libraries. 
Different symbols correspond to our {\small ATLAS9 grid} (blue diamonds), the WLBC99 (green triangles), and 
the G\"ottingen (red circles) spectral libraries. Two solar metallicity isochrones for 20~Myr and 14~Gyr are also shown.\label{fig:bc}}
\end{center}
\end{figure}

The ATLAS9 grid of spectra is complemented by two addiitonal spectral libraries, to cover the parameter space of
cool giants and low-mass dwarfs. At low ${\rm T_{eff}}$ and 
surface gravities, we use the BaSeL WLBC99 results \citep{Westera99, Westera02}. 
This is a semi-empirical library, built from a grid of theoretical spectra that have been later 
calibrated to match empirical color-$T_{eff}$ relations from neighborhood stars. These templates
are available in the metallicity 
range ${\rm -2.0 \le [Fe/H] \le 0.5}$, 
in steps of 0.5~dex. For the low $T_{eff}$ and high gravity regime, we use spectra from the G\"ottingen Spectral Library \citep{Husser13}. These have been 
calculated using the code {\small PHOENIX} \citep{Hauschildt99}, which is particularly suited to model atmospheres 
of cool dwarfs. The {\small PHOENIX} configuration used for this library employs a variable parametrization of microturbulence and mixing length, depending on 
the properties of the modelled atmosphere. The metallicity coverage is ${\rm -4.0 \le [Fe/H] \le 1.0}$, in steps of 0.5~dex. Figure~\ref{fig:bc} shows 
the range of effective temperature and surface gravity covered by our adopted spectral libraries.

We have computed tables of bolometric corrections (BCs) for several popular photometric 
systems (the complete list is found in Table~\ref{tab:bc}), following the prescription by \citet{girardi02} for photon-counting defined systems:

\begin{eqnarray}
BC_{S_\lambda} =  M_{bol,\odot} - 2.5 \log{\Bigl[4\pi (10pc)^2 F_{bol}/L_{\odot}\Bigr]} \nonumber \\
+ 2.5 \log{\Biggl(\frac{\int_{\lambda_1}^{\lambda_2}\lambda F_\lambda S_\lambda d\lambda}{\int_{\lambda_1}^{\lambda_2} \lambda f^0_\lambda S_\lambda d\lambda}\Biggr)} - m^0_{S_\lambda}
\label{eq:bc}
\end{eqnarray}

where $S_\lambda$ is a generic filter response curve, defined between $\lambda_1$ and $\lambda_2$, $F_{bol}=\sigma T_{eff}^4$ is the total emerging flux at the stellar 
surface, $F_\lambda$ is the stellar emerging flux at a given wavelength, $f^0_\lambda$ is the wavelength-dependent flux of a reference spectrum and $m^0_{S_\lambda}$ is the 
magnitude of the reference spectrum in the filter $S_\lambda$ (denoted as zero point). We adopt $M_{bol,\odot} = 4.74$, following the IAU B2 resolution of 2015 \citep{Mamajek15}.

The reference spectra are either the spectrum of Vega ($\alpha \ Lyr$), for 
systems that use Vega for the magnitude zero points (Vegamag systems), or a spectrum 
with constant flux density per unit frequency $f^0_\nu = 3.631 \cdot 10^{-20} {\rm erg\, s^{-1}\, cm^{-2}\, Hz^{-1}}$, for ABmag systems. 
For older photometric systems, such as 
the Johson-Cousins-Glass UBVRIJHKLM we use the energy-integration equivalent of Eq.~\ref{eq:bc}. 

\begin{figure}[ht!]
%\plotone{bolometric-join-criteria.eps}
\begin{center}
\includegraphics[width=3.4in]{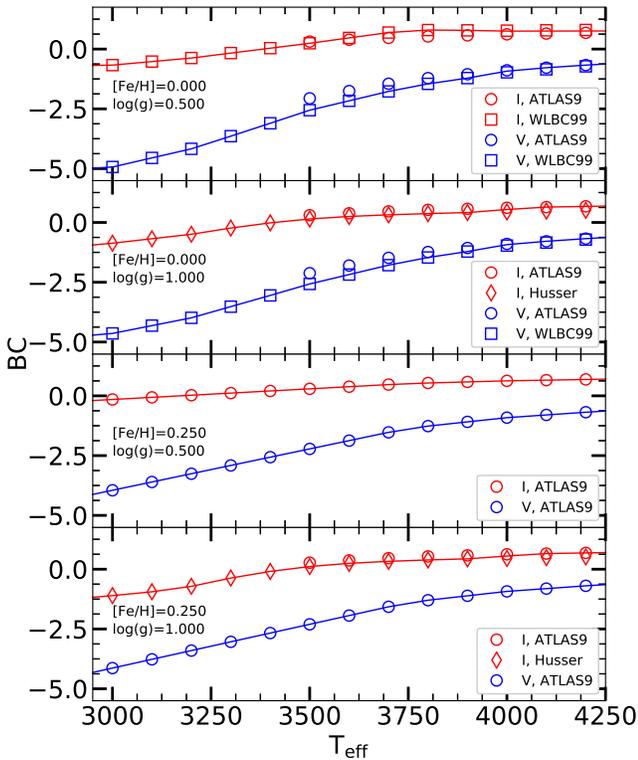}
\caption{An example of  our final BC set (solid lines) for the V and I photometric passbands, 
as a function of the effective temperature, for some selected metallicities 
and surface gravities (see text for details).\label{fig:matchbc}}
\end{center}
\end{figure}

\begin{deluxetable*}{l|ccc}
\tablecolumns{4} 
\tablewidth{0pc}  
\tablecaption{Available photometric systems. We also list the source for the passband definitions and reference zero-points.\label{tab:bc}}
\tablehead{ 
\colhead{Photometric system} & \colhead{Calibration} &  \colhead{Passbands} & \colhead{Zero-points}}
\startdata 
\hline   
UBVRIJHKLM & Vegamag & \citet{Bessel88,Bessel90}  &\citet{Bessel98}\\
HST - WFPC2 & Vegamag &  SYNPHOT &SYNPHOT\\
HST - WFC3 & Vegamag &  SYNPHOT &SYNPHOT\\
HST - ACS & Vegamag & SYNPHOT  &SYNPHOT\\
2MASS & Vegamag & \citet{Cohen03}  &\citet{Cohen03}\\
DECam & ABmag & DES collaboration  & 0\\
Gaia & Vegamag & \citet{Jordi10}\tablenotemark{a}  &\citet{Jordi10}\\
JWST - NIRCam & Vegamag & JWST User Documentation\tablenotemark{b}  &SYNPHOT\\
SAGE & ABmag & SAGE collaboration\tablenotemark{c}   & 0\\
Skymapper & ABmag & \citet{bbs}  & 0\\
Sloan & ABmag &  \citet{Fukugita96} &\citet{Dotter08}\\
Str\"{o}mgren & Vegamag &\citet{Maiz-Apellaniz06}   &\citet{Maiz-Apellaniz06}\\
VISTA & Vegamag &  ESO &\citet{vistazero}\\
\hline
\enddata 
\tablenotetext{a}{The nominal G passband curve has been corrected following the post-DR1 correction provided by \citet{gband}. }
\tablenotetext{b}{https://jwst-docs.stsci.edu/}
\tablenotetext{c}{Zan et al. 2017, Progress in Astronomy, submitted to.}
\end{deluxetable*}

Due to the differences between the adopted sets of spectral libraries, 
the resulting BCs display non-negligible differences in the overlapping ${\rm T_{eff}}$ and
surface gravity regimes. To eliminate discontinuities in the final merged BC set, 
the different sets were matched smoothly in the overlapping regions by applying some 
suitable ramping at the edge of the various tables. After several tests we adopted the following combination of BC libraries: 

\begin{itemize}

\item{at metallicities equal or lower than solar, for the V passband (or passbands with equivalent effective wavelengths) and 
all photometric passbands bluer than the V-band we employ the BCs from our   
{\small ATLAS9} grid, supplemented at lower gravities and $T_{eff} < 3900$~K by WLBC99 results.  
For redder passbands and $T_{eff} < 3900$~K we switch at ${\rm \log(g)=1.0}$ from our {\small ATLAS9} BCs  
to \cite{Husser13} BCs for higher gravities, and to WLBC99 BCs for lower gravities;}

\item{at super solar metallicities, we adopt our {\small ATLAS9} BCs for the V band (or equivalent) as well as for bluer photometric passbands, extrapolating 
linearly in log(g) and $T_{eff}$ when necessary. For redder photometric 
passbands we use {\small ATLAS9} BCs for gravities lower than ${\rm \log(g)=1.0}$ (extrapolated when necessary) 
and \cite{Husser13} BCs for gravities larger or equal than this limit, and $T_{eff} < 3900$~K.}

\end{itemize}

Figure~\ref{fig:matchbc} shows examples of our adopted composite BC library.

\subsection{Asteroseismic properties of the models} \label{sec:seismic}

Asteroseismology has experienced a revolution thanks to past and present space missions such as CoRoT \citep{bagl09}, {\it Kepler} \citep{gill10b}, and K2 \citep{chap15}, which have provided high-precision photometric data for hundreds of main-sequence and sub-giant stars and for thousands of red giants. 

Future satellites like TESS \citep{rick14} and PLATO \citep{raue14} hold promises to expand the current sample greatly and thus further extend the impact of asteroseismology in the fields of stellar physics \citep[e.g.,][]{Beck:2011jr,Verma:2014fx}, exoplanet studies \citep[e.g.,]{Huber:2013jb,2015MNRAS.452.2127S}, and Galactic archaeology \citep[e.g.,][]{2016MNRAS.455..987C,2017arXiv171009847S}. Given the availability of high-quality oscillations data, we provide the corresponding theoretical quantities to fully exploit their potential.

We have computed adiabatic oscillation frequencies for all the models using the Aarhus aDIabatic PuLSation package \citep[ADIPLS,][]{jcd08a}. We provide the radial, dipole, quadrupole, and octupole mode frequencies for the models with central hydrogen mass fraction $>10^{-4}$ and only the radial mode frequencies for more evolved models. {\bf The power spectrum of the solar-like oscillators have several global characteristic features that can be used to constrain the stellar properties. Some of these features do not require very high signal-to-noise data for their determinations --in contrast to the individual oscillation frequencies which need long time-series data with high signal-to-noise ratio for their measurements - and play a crucial role in ensemble studies.}  We also provide three {\bf such} global asteroseismic quantities for the models, viz., the frequency of maximum power ($\nu_{\rm max}$), large frequency separation for the radial mode frequencies ($\Delta\nu_0$), and the asymptotic period spacing for the dipole mode frequencies ($\Delta P_1$).

\begin{figure}
\epsscale{1.2}
\plotone{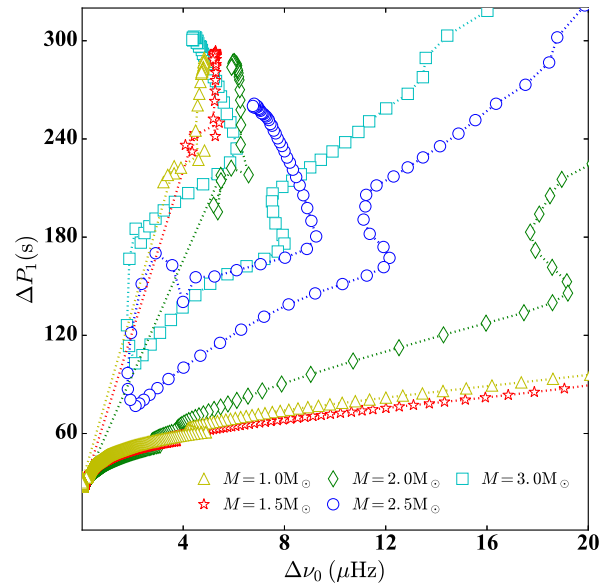}
\caption{Asymptotic period spacing as a function of the large frequency separation for a set of 5 tracks 
with different masses and teh same initial composition ($Y = 0.26$ and $[{\rm Fe}/{\rm H}] = -0.2$~dex). \label{period_spacing}}
\end{figure}

The value of $\nu_{\rm max}$ was determined using the well known scaling relation \citep{kjel95},
\begin{equation}
\frac{\nu_{\rm max}}{\nu_{{\rm max},\odot}} = \left(\frac{M}{M_\odot}\right)\left(\frac{R}{R_\odot}\right)^{-2}\left(\frac{T_{\rm eff}}{T_{{\rm eff},\odot}}\right)^{-1/2},
\end{equation}
where $M$, $R$, and $T_{\rm eff}$ are the model mass, radius, and effective temperature, respectively. We adopted $\nu_{{\rm max},\odot}=3090~\mu$Hz from \citet{hube11}, $T_{{\rm eff},\odot} = 5777$~K, and $M_\odot = 1.9891\times10^{33}$~gm and $R_\odot = 6.9599\times10^{10}$~cm as used in the corresponding stellar tracks. We extracted $\Delta\nu_0$ following \citet{whit11}, i.e., performing a weighted linear least squares fit to the radial mode frequencies as a function of the radial order, with a Gaussian weighting function centered around $\nu_{\rm max}$, with $0.25\,\nu_{\rm max}$ full width at half maximum. {\bf The large frequency separation and frequency of maximum power, together with the measurement of the stellar $T_{eff}$, have been used to determine masses and radii of large samples of isolated stars, independent of modelling, thus providing strong constraints on stellar evolution models and on models of Galactic stellar populations \citep[see, e.g.,][]{kall10,chap11,migl12}.}

We determined the period spacing $\Delta P_1$ using the asymptotic expression,
\begin{equation}
\Delta P_1 = \sqrt{2}\pi^2 \left(\int \frac{N}{r} dr \right)^{-1},
\end{equation}
where $N$ and $r$ are the Brunt-V\"{a}is\"{a}l\"{a} frequency and radial coordinate, respectively. The integration is performed over the radiative interior. {\bf Since $N$ is weighted with $r^{-1}$ in the integral, $\Delta P_1$ is very sensitive to the Brunt-V\"{a}is\"{a}l\"{a} frequency profile in the core. Hence the measurement of $\Delta P_1$ offers a unique opportunity to constrain the uncertain aspects of the physical processes taking place in stellar cores. As an example, \citet{degr10} used the measurement of the period spacing for the star HD 50230 
observed using the CoRoT satellite, to constrain the mixing in its core \citep[see also,][]{mont13}.} Figure~\ref{period_spacing} illustrates the evolution of models in the {\bf $\Delta\nu_0-\Delta P_1$} diagram (evolution proceeds from right to left). {\bf This is an interesting diagram because $\Delta\nu_0$ contains information mostly about the envelope, whereas $\Delta P_1$ about the core.} The hook-like feature on the right (beyond the displayed range for ${\rm M =1.0}$ and 1.5~${\rm M_\odot}$) correspond to the base of the red giant branch. The sudden jump at the lowest $\Delta\nu_0$ for ${\rm M =}$ 1.0, 1.5, and 2.0~${\rm M_\odot}$ is due to the helium flash, which causes the stellar structure to change rapidly in a short period of time. This diagram have been used successfully to distinguish the shell hydrogen burning red giant stars with those that are fusing helium in the core 
along with the hydrogen in the shell {\bf \citep[e.g.,][]{bedd11,moss11}}.

\section{Comparisons with existing model databases}
\label{databases}

This section is devoted to comparisons of our isochrones with recent, widely employed 
isochrone and stellar model databases. The goal is to give a general picture of how our new 
calculations compare to recent, popular models. The model grids shown in our comparisons  
are computed employing various different choices for the input physics and treatment of mixing, 
and also the reference solar metal distribution can be different (see Tables~\ref{differencesa} and ~\ref{differencesb} for a summary). 
We show comparisons in the HRD, to bypass the additional degree of freedom introduced by the 
choice of the bolometric corrections.

\begin{figure}[b]
% \vspace*{-2.0 cm}
\begin{center}
 \includegraphics[width=3.2in]{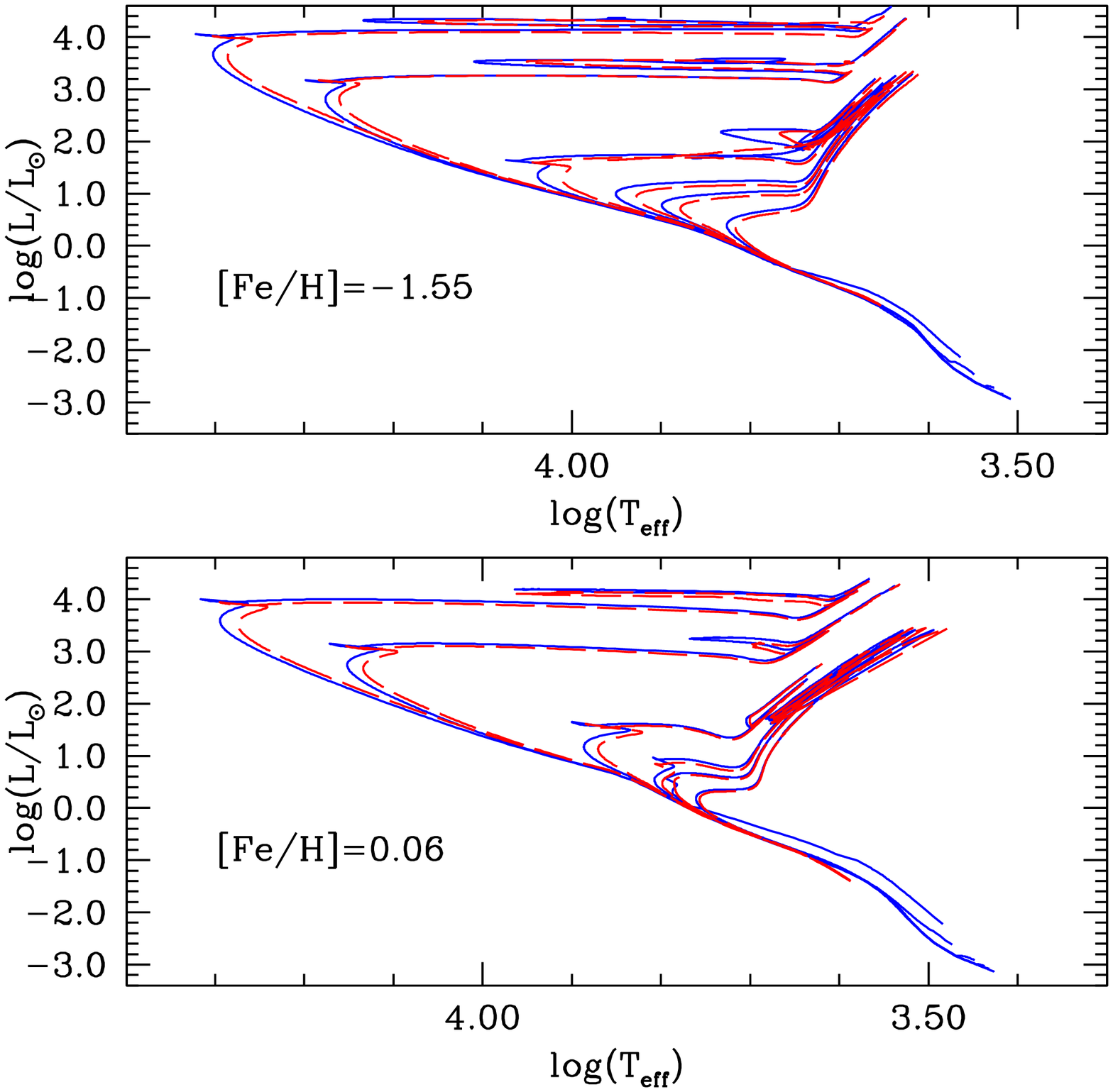} 
% \vspace*{-1.0 cm}
 \caption{Comparison of our isochrones for [Fe/H]=0.06 and [Fe/H]=$-$1.55 (solid lines) with the 
older BaSTI isochrones for [Fe/H]=0.06 and [Fe/H]=$-$1.49 (dashed lines), and ages equal to 
30~Myr, 100~Myr, 1~Gyr, 3~Gyr, 5~Gyr and 12~Gyr, respectively (see text for details).}
   \label{figurebastiold}
\end{center}
\end{figure}

We start first with a comparison with our previous BaSTI computations \citep{basti:04}, displayed in Fig.~\ref{figurebastiold}. We show 
our new isochrones for (Fe/H]=0.06 and [Fe/H]=$-$1.55,  
and ages equal to 30~Myr, 100~Myr, 1~Gyr, 3~Gyr, 5~Gyr and 12~Gyr, respectively, compared 
to the older BaSTI release for the same ages, [Fe/H]=0.06 and [Fe/H]=$-$1.49 (the metallicity grid point closest to [Fe/H]=$-$1.55 
in the older release) and $\eta$=0.4. 
We consider here our new isochrones without diffusion, because the older model grid was calculated neglecting atomic diffusion 
{\bf (we are using our set b) of models as described in Table~\ref{tab:grids})}. Core overshooting 
during the MS is included in both sets of isochrones.
Notice that the total metal mass fraction $Z$ is lower in the new 
isochrones, due to the different solar heavy element distribution.

\begin{figure}[b]
% \vspace*{-2.0 cm}
\begin{center}
 \includegraphics[width=3.4in]{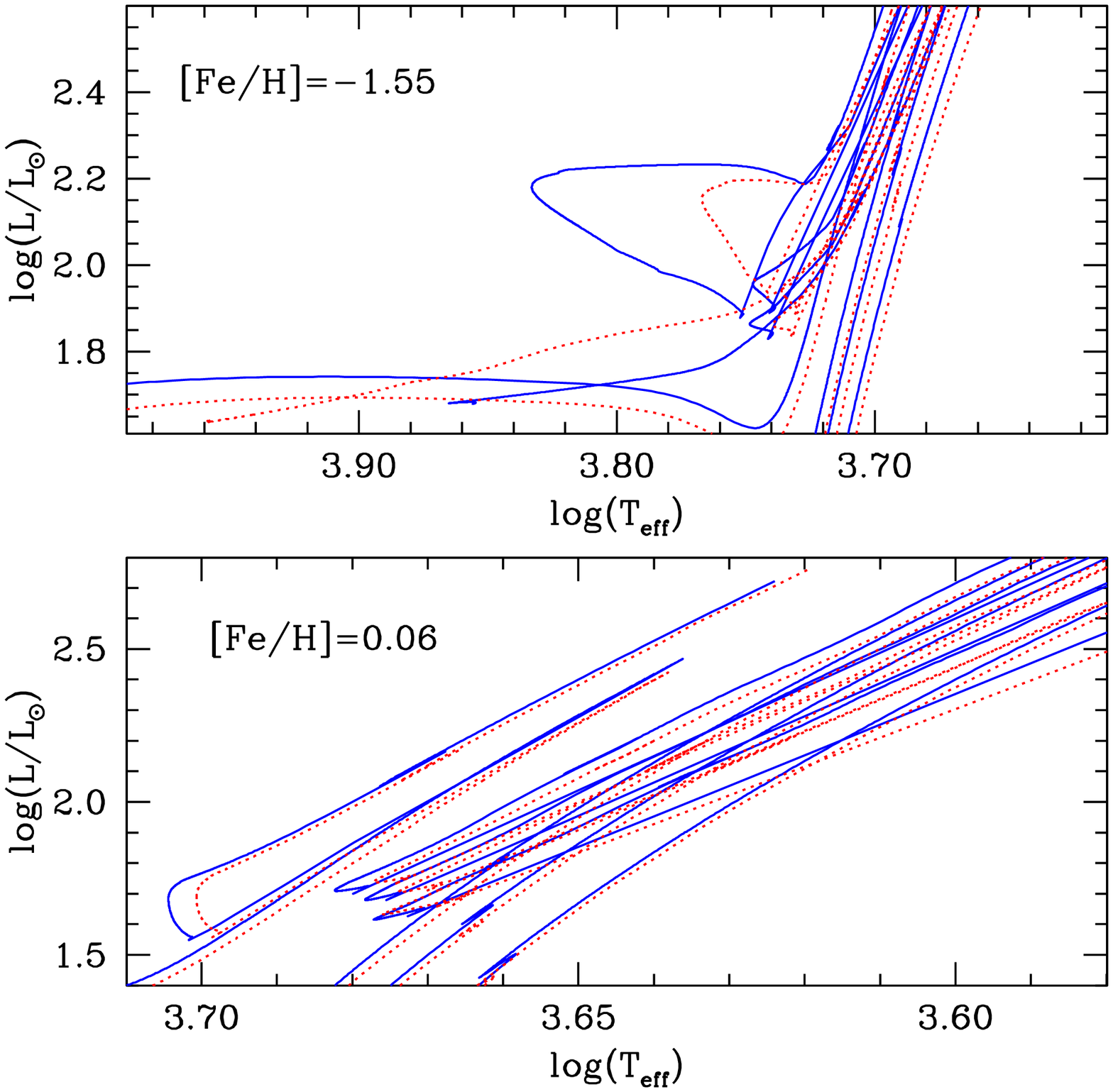} 
% \vspace*{-1.0 cm}
 \caption{As Fig.~\ref{figurebastiold} but showing the core He-burning region for ages between 1 and 12~Gyr.
The older BaSTI isochrones are displayed as dotted lines.}
   \label{figbastioldHB}
\end{center}
\end{figure}

The new isochrones have slightly hotter RGBs, and TO. The core He-burning sequences are brighter 
for ages below 1~Gyr, and the HRD blue-loops are generally more extended.
Figure~\ref{figbastioldHB} enlarges the core He-burning portion of the isochrones for 
ages between 1 and 12~Gyr. The new isochrones have slightly fainter luminosities  (by a few hundredth dex) 
during core He-burning at these ages --mainly because of the new electron conduction opacities--
and slightly hotter effective temperatures, as for the RGB.
At 12~Gyr and [Fe/H]$-$1.55 the new isochrones show a cooler He-burning phase, 
because of the lower of $\eta$ used in the new calculations. 

{\bf The main reason for the differences between these new BaSTI computations and the previous ones is the updated  
solar metal distribution and associated lower $Z$ at a given [Fe/H]. However, the lower luminosity of the core He-burining phase 
at old ages is driven by the updated electron conduction opacities employed in these new calculations. 
}

%\resizebox{\linewidth}{!}{
%
%}

%\begin{rotatetable}
\begin{deluxetable*}{lcccc}
\tablecaption{Main differences amongst the physics inputs and solar metal mixture adopted in our calculations and the 
independent calculations discussed in this section. The symbol \lq{---}\rq denotes the same treatment as in our calculations. 
\label{differencesa}}
\tablewidth{700pt}
\tabletypesize{\scriptsize}
\tablehead{
  \colhead{Code}  & \colhead{EOS} & \colhead{Reaction rates} & \colhead{Opacity} & \colhead{Solar mix} }
\startdata
\cite{tpd11}  & OPAL                 & ---& ---&  \cite{asp05}\\ 
(pre-MS)      & \citep{rn02}         &     &     &              \\
\\
\hline
\\
\cite{sdf00}  & Own calculations               & \cite{cf88} & low-T opacities \citep{af}  & \cite{gn93}\\ 
(pre-MS)      &                                &             & electron conduction \citep{iben} &            \\  
\\
\hline
\\
PARSEC        & --- & JINA REACLIB   & low-T opacities \citep{aeso} & -- \\ 
              &    & \citep{cyburt} & electron conduction \citep{itoh}  &    \\  
\\
\hline
\\
MESA          &\cite{sc95} & JINA REACLIB  & --- &  \cite{asp09} \\ 
              &\cite{rn02} &   &    & \\  
              &\cite{mm12} &   &    &  \\  
\enddata
\end{deluxetable*}
%\end{rotatetable}

%\begin{rotatetable}
\begin{deluxetable*}{lcccc}
  \tablecaption{As Table~\ref{differencesa}, but for the differences in the treatment of convective mixing,
    mass loss, mixing length and outer boundary conditions.  
\label{differencesb}}
\tablewidth{700pt}
\tabletypesize{\scriptsize}
\tablehead{
  \colhead{Code}  & \colhead{Mixing} & \colhead{Reimers $\eta$ and $\alpha_{\rm ML}$}   &  \colhead{Bound. cond.}  &  \colhead{Diffusion} }
\startdata
\cite{tpd11}  & --- &  $\eta$=0.0  &  theoretical        &            \\
(pre-MS)      &    & $\alpha_{\rm ML}$=1.9   &  model atmospheres  &    \\
\\
\hline
\\
\cite{sdf00}  & --- &  $\eta$=0.0  &  theoretical          & --- \\
(pre-MS)      &    &  $\alpha_{\rm ML}$=1.605   &  model atmospheres    &   \\
\\
\hline
\\
PARSEC        & proportional mean free path {\sl across} &  $\eta$=0.2 &  gray $T(\tau)$ plus {\sl calibrated} &off when conv. envelope \\
              & border all conv. regions \citep{breov}   &  $\alpha_{\rm ML}$=1.74          & $T(\tau)$ for VLM models & mass below a threshold  \\ 
\\
\hline
\\
MESA          & Ledoux criterion, diffusive mixing         &  $\eta$=0.1 (RGB) &  theoretical       & moderated  with \\
              & diffusive overshooting/semiconv.           &  $\eta$=0.2 (AGB) &  model atmospheres & diffusive mixing \\
              &  &   $\alpha_{\rm ML}$=1.82   &   & \\
              &  &   \citep{henyey} formalism & & \\
\enddata
\end{deluxetable*}
%\end{rotatetable}

\subsection{Pre-MS isochrones}

We have compared our new isochrones with independent calculations, considering   
separately pre-MS isochrones for low- and very low-mass stars, that with our grid of models can be calculated 
for a minimum age of just 4~Myr, whereas complete isochrones reaching the AGB phase or C-ignition 
start from an age of 20~Myr.

\begin{figure}[b]
% \vspace*{-2.0 cm}
\begin{center}
 \includegraphics[width=3.2in]{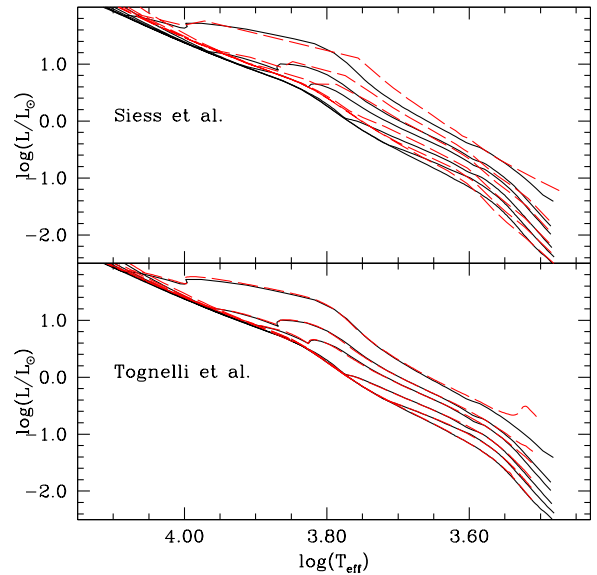} 
% \vspace*{-1.0 cm}
 \caption{Comparison of our pre-MS isochrones (solid lines) with \cite{sdf00} and \cite{tpd11} results (dashed lines in the 
top and bottom panel, respectively) for a metallicity around solar 
and ages equal to  4, 10, 15, 30, 50, and 100~Myr, respectively (see text for details).}
   \label{figurepreMS}
\end{center}
\end{figure}

The pre-MS isochrones have been compared to results from 
the extensive database by \cite{tpd11}, and the \lq{classic\rq} models by \cite{sdf00}, 
as shown in Fig.~\ref{figurepreMS}. 
These latter two calculations differ from ours concerning some physics inputs. In particular, 
\cite{tpd11} isochrones have been calculated adopting a different EOS and boundary conditions, 
whilst \cite{sdf00} isochrones have been computed with different low-temperature radiative opacities, 
EOS, boundary conditions, and the initial deuterium abundance is about half the value used in our calculations. 
The reference solar metal mixture is different for each of the three 
sets of isochrones shown in the figure.
The minimum evolving mass along the isochrones is equal to 0.1$M_{\odot}$ for our and \cite{sdf00} calculations, 
while it is equal to  0.2$M_{\odot}$ for \cite{tpd11} models.

For the comparison we have selected \cite{tpd11} calculations (that at fixed $Z$ allow for various choices of $Y$, 
the deuterium mass fraction $X_D$ and mixing length) for 
$Z$=0.0175, $Y$=0.265 $X_D=4 \ 10^{-5}$, $\alpha_{\rm ML}$=1.9 --very close to our
initial solar chemical composition, the adopted 
initial deuterium mass fraction and solar calibrated mixing length-- and the $Z$=0.02 \cite{sdf00} isochrones.
We have considered ages equal to 4, 10, 15, 30, 50, and 100~Myr, respectively. The upper age limit is fixed by the 
largest age available for \cite{tpd11} calculations.
 
The agreement between our $Z$=0.0172 ([Fe/H]=0.06) and \cite{tpd11} isochrones is remarkable. They are almost indistinguishable, 
appreciable differences appearing only for the lowest masses 
in common and the two youngest ages, where \cite{tpd11} isochrones are more luminous than ours at a given $T_{eff}$. 
Differences with respect to \cite{sdf00} calculations are larger and more systematic, their isochrones being almost always 
brighter at fixed $T_{eff}$ for stellar masses between $\sim$2.0-2.5 $M_{\odot}$ and $\sim$0.4$M_{\odot}$.

\subsection{MS and post-MS isochrones}

Our complete isochrones have been compared with results from the recent PARSEC and MIST isochrones.
We considered the non-rotating MIST isochrones, and the PARSEC isochrones with VLM stellar models calculated with 
the \lq{calibrated}\rq\ boundary conditions, as described in \cite{chen14}.

We considered our isochrones including convective core overshooting during 
the MS and atomic diffusion {\bf (the reference set a) described in Table~\ref{tab:grids})}, for both effects are included in the MIST and PARSEC isochrones, 
although with varying implementations.  
Compared to our models, the non-rotating MIST isochrones have been calculated with 
different implementations of convective mixing (and include thermohaline 
mixing during the RGB), as well as different choices for the solar metal distribution, 
EOS, reaction rates, boundary conditions, mixing length theory formalism, 
and a lower value of the Reimers $\eta$ parameter. Radiative levitation 
is neglected, and the efficiency of atomic diffusion during the MS is moderated by including a competing turbulent diffusive 
coefficient \citep[see][for details]{mist}.

The PARSEC calculations have employed, compared to our new models, different choices for the 
low-temperature radiative opacities, electron conduction opacities, reaction rates, 
implementation of overshooting, boundary conditions, and a lower value of the Reimers parameter $\eta$. 
Atomic diffusion without radiative levitation is included, 
but switched off when the mass size of the outer convective region decreases below a given threshold 
\citep[see][for details]{parsec}.

\begin{figure}[b]
% \vspace*{-2.0 cm}
\begin{center}
 \includegraphics[width=3.2in]{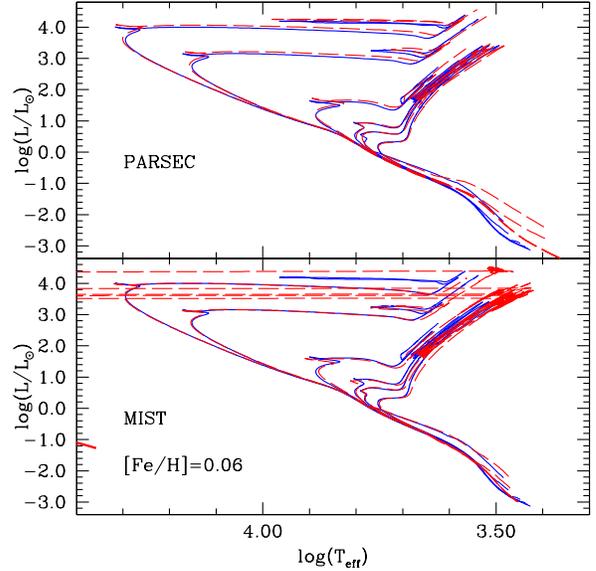} 
% \vspace*{-1.0 cm}
 \caption{Comparison of our complete isochrones for [Fe/H]=0.06 (solid lines) with PARSEC and MIST results (dashed lines in the 
top and bottom panel, respectively) and ages equal to 
30~Myr, 100~Myr, 1~Gyr, 3~Gyr, 5~Gyr and 12~Gyr, respectively (see text for details).}
   \label{figp06complete}
\end{center}
\end{figure}

\begin{figure}[b]
% \vspace*{-2.0 cm}
\begin{center}
 \includegraphics[width=3.2in]{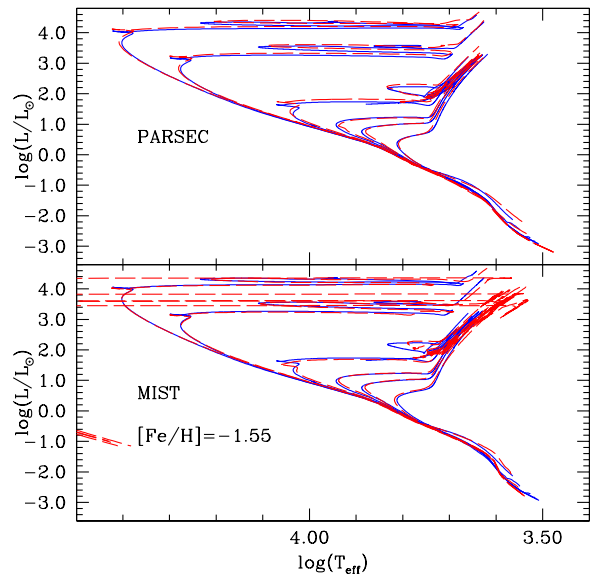} 
% \vspace*{-1.0 cm}
 \caption{As Fig.~\ref{figp06complete} but for [Fe/H]=$-$1.55 (see text for details).}
   \label{figm155complete}
\end{center}
\end{figure}

Figures~\ref{figp06complete} and \ref{figm155complete} show selected isochrones for 30~Myr, 100~Myr, 1~Gyr, 
3~Gyr, 5~Gyr and 12~Gyr, [Fe/H]=0.06 and [Fe/H]=$-$1.55, respectively. 
They are shown together with PARSEC isochrones for the same ages, [Fe/H]=0.07 and $-$1.59\footnote{Retrieved with the 
web interface at \url{http://stev.oapd.inaf.it/cgi-bin/cmd}}, and MIST 
isochrones for the same ages and [Fe/H] of our isochrones 
\footnote{Retrieved with the MIST web interpolator at \url{http://waps.cfa.harvard.edu/MIST/interp_isos.html}}.

\begin{figure}[b]
% \vspace*{-2.0 cm}
\begin{center}
 \includegraphics[width=3.2in]{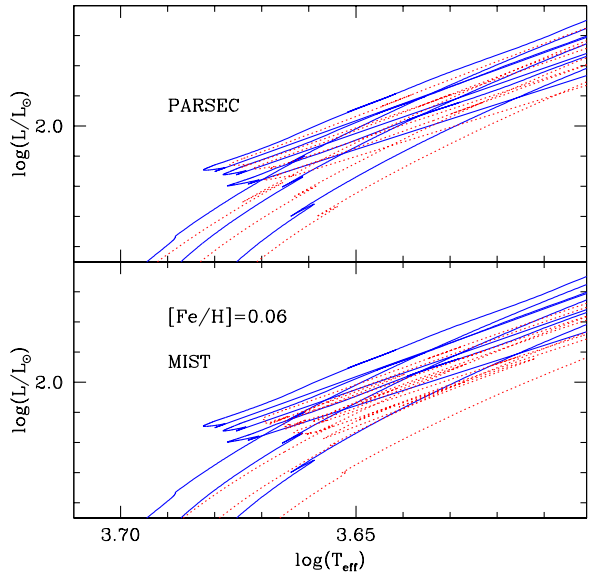} 
% \vspace*{-1.0 cm}
 \caption{As Fig.~\ref{figp06complete} but showing the core He-burning region for ages between 1 and 12~Gyr.
MIST and PARSEC isochrones are displayed as dotted lines.}
   \label{figp06He}
\end{center}
\end{figure}

\begin{figure}[b]
% \vspace*{-2.0 cm}
\begin{center}
 \includegraphics[width=3.2in]{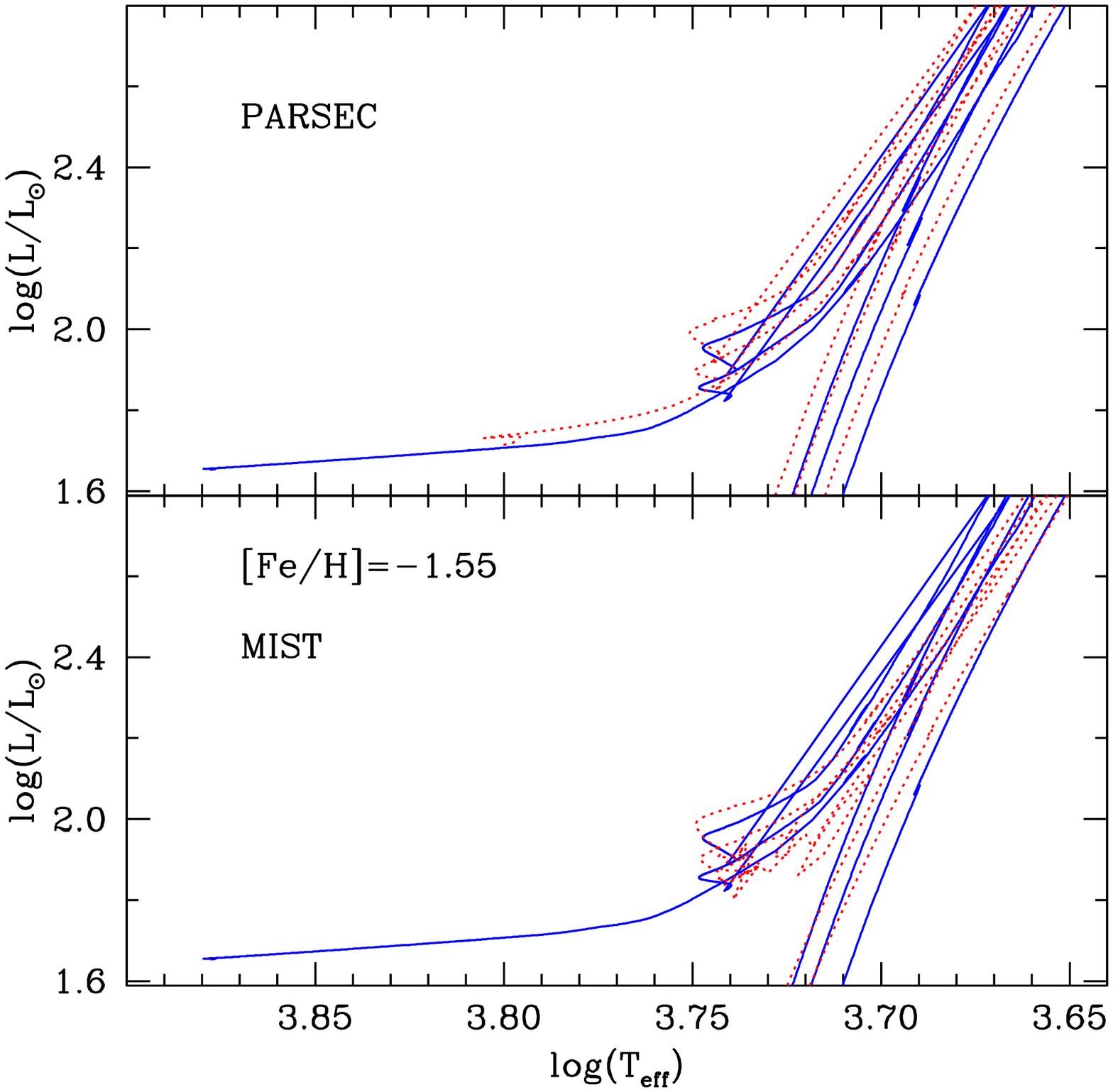} 
% \vspace*{-1.0 cm}
 \caption{As Fig.~\ref{figp06He} but for [Fe/H]=$-$1.55.}
   \label{figm155He}
\end{center}
\end{figure}

The comparison with PARSEC isochrones displays a remarkable general agreement especially 
at the lower [Fe/H], whereas at the higher metallicity the lower masses (that are still evolving 
along the pre-MS phase in the youngest two isochrones) are systematically discrepant compared to our models. 
The TO luminosities are only slightly different, especially at the three lowest ages, where 
the effect of different core overshooting prescription may play a role.
The core He-burning phase is slightly over-luminous compared to our models, RGBs 
are slightly cooler compared to our [Fe/H]=0.06 isochrones, and slightly hotter compared to the 
[Fe/H]=$-$1.55 ones.
Figures~\ref{figp06He} and \ref{figm155He} enlarge the core He-burning portion of the isochrones for 
ages between 1 and 12~Gyr. The RGB of PARSEC isochrones is cooler by less than 100~K 
compared to our models for [Fe/H]=0.06, and hotter by less than 100~K at the lower metallicity.
The luminosity of the He-burning phase is only slightly larger (by a few hundredth dex) at both metallicities.
Notice that at 12~Gyr the start of quiescent core He-burning in our isochrones 
is at a hotter $T_{eff}$ than PARSEC results, 
due to our choice of a larger $\eta$ Reimers parameter.

The comparison with MIST isochrones yields similar results. There is an overall good agreement for the 
MS, TO, subgiant-branch (SGB) phases, and also in the regime of the lowest masses, still evolving along the pre-MS at the youngest ages. 
The He-burning phase of MIST isochrones is generally over-luminous, RGBs systematically redder at [Fe/H]=0.06, and with a 
different slope at [Fe/H]=$-$1.55.
Figures~\ref{figp06He} and \ref{figm155He} show RGBs over 100~K cooler than our models at [Fe/H]=0.06, 
and slightly larger core He-burning luminosities, like in the comparison with PARSEC. Also 
in comparison with MIST isochrones, at 12~Gyr the start of quiescent core He-burning in our isochrones 
is at a hotter $T_{eff}$, again due to our choice of a larger $\eta$ Reimers parameter.
 
\section{Comparisons with data}
\label{obs}

In this section we present results of some tests, performed 
to assess the general consistency of our new models and isochrones with constraints coming 
from eclipsing binary analyses, stars with asteroseismic mass determinations, and star clusters. 
The isochrones used in these comparisons include convective core overshooting during the MS for the 
appropriate age range and neglect atomic diffusion during the MS {\bf (set b) of models described in Table~\ref{tab:grids})}, 
if not otherwise specified. 

\subsection{Binaries}

We first consider masses and radii for 
pre-MS detached eclipsing binary (DEB) systems compiled by \cite{stassun} 
and \cite{sk17}, covering a mass range between 0.2 and 4.0$M_{\odot}$. We assume an initial [Fe/H]=0.06 
\citep[overall consistent with the few available spectroscopic estimates, see][]{stassun}, and 
consider a minimum age of 4~Myr, the lowest possible value with our model grid. 
We do not aim to find a best-fit solution for all the systems, 
just at least one isochrone that matches simultaneously mass and radius of 
both components for each system within the errors, to denote a general consistency between models and observations. 

{\bf This test is relevant for the general adequacy of both boundary conditions and $\alpha_{\rm ML}$ value 
employed in the calculations, given the extreme sensitivity of pre-MS tracks to the combination of these two inputs. It is however worth 
noticing that the lack of model-independent age estimates prevent this type of tests from providing very stringent constraints on the models.}

\begin{figure}[b]
% \vspace*{-2.0 cm}
\begin{center}
 \includegraphics[width=3.2in]{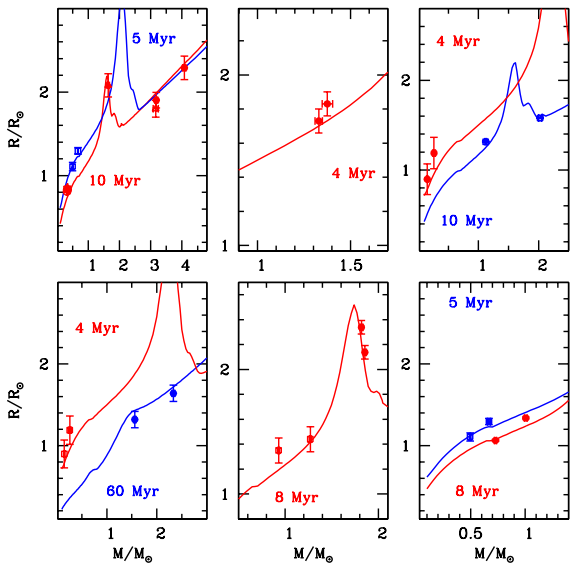} 
% \vspace*{-1.0 cm}
 \caption{Comparison in the MR diagram between our pre-MS isochrones and a sample of pre-MS 
DEB systems (see text for details). Notice the local maximum of the radius displayed by the 
4, 5, 8, and 10~Myr isochrones, corresponding to C and N abundances attaining their equilibrium abundances.}
   \label{figEBprems}
\end{center}
\end{figure}

We found 13 systems in the age range covered by our pre-MS isochrones, displayed in a mass-radius (MR) 
diagram in Fig.~\ref{figEBprems}.
In case of all these systems, our isochrones can match both components within the quoted 1$\sigma$ error bars  
with a single age value, varying between 5 and 60 Myr among the whole sample of DEBs. 

The next test involves low-mass MS models.
It has been recognized since some time the existence of a disagreement between 
theoretical and observational MR relationships for low-mass stars, with model radii 
typically 10-20\% smaller than observations for a fixed mass, \citep[see, e.g.,][for a review]{tag10}.
Here we examine first the level of agreement between the observed and theoretical MS low-mass  
MR relationship, by comparing our grid of models with data 
from DEB systems that host components with M$< 0.8~M_{\odot}$, as compiled by \cite{feiden}. This compilation includes  
systems with quoted random uncertainties in
both mass and radius below 3\%. As for the pre-MS case, the requirement for the models is 
that they are able to match the position  
of both system components in the MR diagram for a single value of the age. 

\begin{figure}[b]
% \vspace*{-2.0 cm}
\begin{center}
 \includegraphics[width=3.2in]{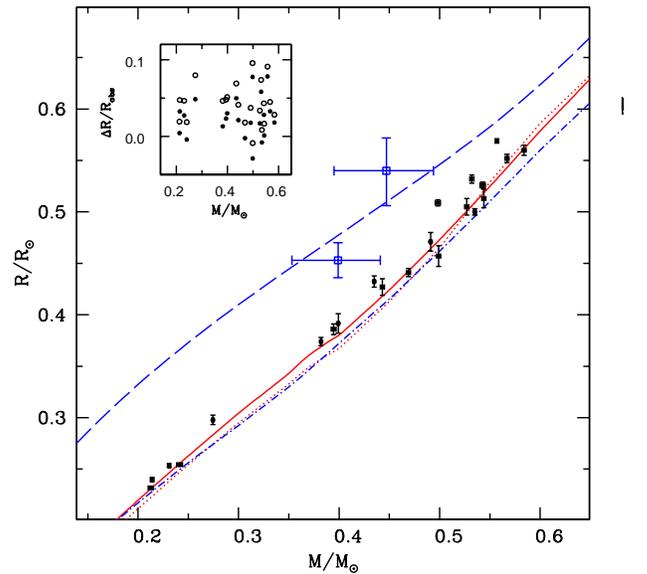} 
% \vspace*{-1.0 cm}
 \caption{Comparison in the MR diagram between our 1~Gyr (dashed dotted line) and 12~Gyr (solid line) [Fe/H]=0.06 isochrones,    
and a subsample of \cite{feiden} DEB systems whose components are found to be 
evolving on the ZAMS. The dotted line denotes a 12~Gyr old [Fe/H]=$-$0.40 
isochrone, to show the almost negligible effect of metallicity variations when the mass is below $\sim$0.7$M_{\odot}$.
Open squares denote the components of the system KELT~J041621-62004. 
The dashed line displays a 50~Myr, [Fe/H]=0.06 pre-MS isochrone. 
The inset shows the run of the relative radius 
differences (observations-theory) $\Delta R / R_{\rm obs}$ with the mass of the systems' components (bar the system KELT~J041621-62004) for 
an age of 1~Gyr (open circles) and 12~Gyr (filled circles -- see text for details).}
    \label{figEBlms_a}
\end{center}
\end{figure}

We assume that all DEBs have metallicity around solar \citep[see also][for spectroscopic metallicity estimates 
for a few of the systems in their compilation]{feiden} 
and split the sample into two subsamples. The first one is made of systems with both components essentially 
on the ZAMS, displayed in Fig.\ref{figEBlms_a}, together with isochrones of ages equal to 1~Gyr and 
12~Gyr respectively, and [Fe/H]=0.06. We also show a 12~Gyr isochrone for [Fe/H]=$-$0.40, to highlight the 
insensitivity of the theoretical MR relationship to metallicity, when the mass is below $\sim$0.7~$M_{\odot}$.   

\begin{figure}[b]
% \vspace*{-2.0 cm}
\begin{center}
 \includegraphics[width=3.2in]{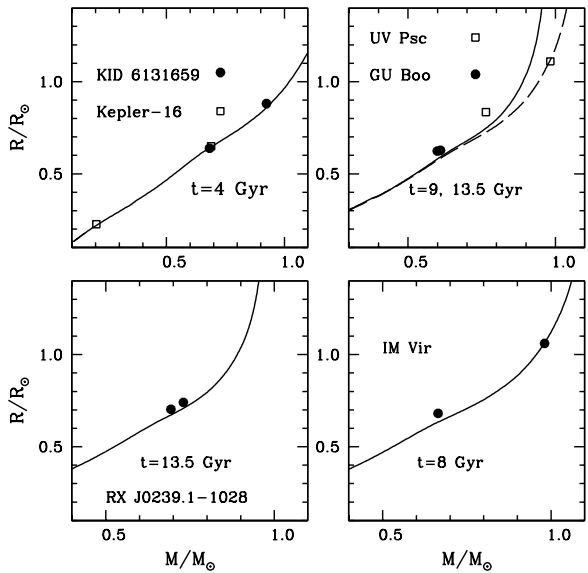} 
% \vspace*{-1.0 cm}
 \caption{As Fig.~\ref{figEBlms_a} but for \cite{feiden} DEB systems with at least one component 
evolved off the ZAMS. }
   \label{figEBlms_b}
\end{center}
\end{figure}

Isochrones appear to match reasonably well all systems, without a clear 
major systematic discrepancy in radius at fixed mass. 
The effect of age is very small for this mass range. If we denote with $\Delta R$ the difference 
$R_{\rm obs}-R_{\rm theory}$ between observed and predicted radius for an object with observed mass $M$, we find 
an average $\Delta R / R_{\rm obs}$=0.02$\pm$0.03 assuming an age of 12~Gyr for all systems, and an average 
$\Delta R / R_{\rm obs}$=0.04$\pm$0.03 for an age of 1~Gyr (see inset of Fig.~\ref{figEBlms_a}). 
These average differences are consistent with typical systematic errors 
on empirical radius estimates --of the order of 2-3\%-- as determined by \cite{woe10} in their reanalysis of the DEB system 
Gu Boo.

The second subsample (see Fig.\ref{figEBlms_b}) includes systems with one or both components evolved off the ZAMS. 
We impose an upper limit of 13.5~Gyr to their ages, to match the cosmological constraint.
The major discrepancy here is for UV~Psc, whereby one component is matched by the 9~Gyr isochrone, whereas the less 
massive one appears older than 13.5~Gyr. A minor discrepancy affects also IM~Vir, with the lower mass component appearing 
slightly older than the companion.

On the whole there is no major systematic discrepancy between models and observed MR relationships, although 
there are clear mismatches for a few cases, as found also by \cite{feiden} analysis. 
Another example of mismatch is the M-dwarf system (both components with 
masses around $0.4~M_\odot$) KELT~J041621-620046 studied very recently by \cite{lubin17}, and shown in 
Fig.\ref{figEBlms_a}. Our 
isochrones give radii systematically lower than observed for both components \citep[as all other models 
employed by][]{lubin17} by $\sim$20\%. The commonly accepted explanation for these mismatches 
\citep[see, e.g.,][and references therein]{feiden, lubin17} involves  
effects of large-scale magnetic fields that suppress convective motions, and increase 
the total surface coverage of starspots. This causes a reduction in the total energy flux across a given surface 
within the star, forcing the stellar radius to inflate and ensure flux conservation.
{\bf For the sake of comparison we also show in Fig.\ref{figEBlms_a} a 50~Myr, [Fe/H]=0.06 pre-MS isochrone that would 
match within the error bars the position of KELT~J041621-620046 components in the MR diagram, in case these objects were actually 
pre-MS stars.}

\begin{figure}[b]
% \vspace*{-2.0 cm}
\begin{center}
 \includegraphics[width=3.2in]{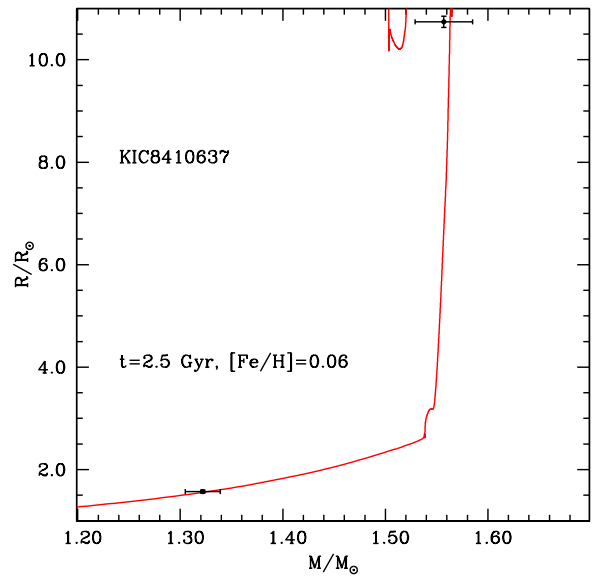} 
% \vspace*{-1.0 cm}
 \caption{As Fig.~\ref{figEBlms_a} but for the components of the KIC 8410637 system (see text for details).}
   \label{figkeplerEB}
\end{center}
\end{figure}

The next comparison involves the DEB system 
KIC 8410637, studied by \cite{frandsen}. It contains a MS and a RGB star, and is another good test for the 
calibration of convection in the models. Figure~\ref{figkeplerEB} compares observations and isochrones in the MR 
diagram. When considering isochrones for [Fe/H]=0.06, consistent with the spectroscopic estimate by 
\cite{frandsen}, we find that an age of 2.5~Gyr matches very well the position of the two components 
\citep[a similar result was found by][using PARSEC isochrones]{frandsen}.
 
\begin{figure}[b]
% \vspace*{-2.0 cm}
\begin{center}
 \includegraphics[width=3.4in]{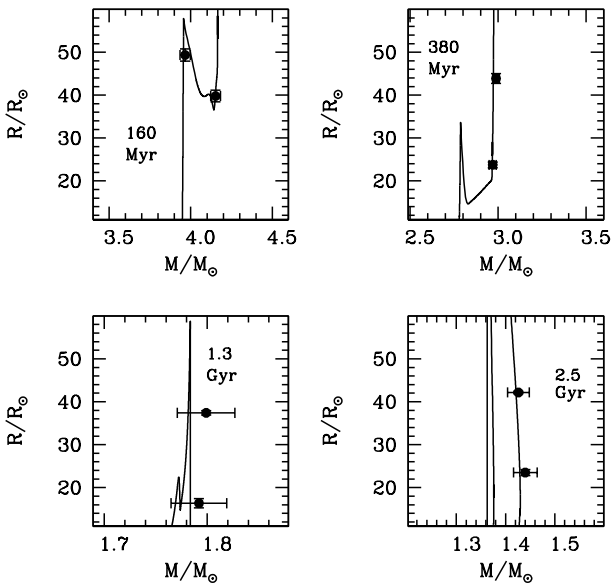} 
% \vspace*{-1.0 cm}
 \caption{As Fig.~\ref{figEBlms_a} but for the components of --moving clockwise from the top left panel--   
OGLE-LMC-ECL-06575, OGLE-LMC-ECL-09660, OGLE-LMC-ECL-15260 
and OGLE-LMC-ECL-03160 systems (see text for details).}
   \label{figClaretEB}
\end{center}
\end{figure}

The last DEB systems compared to our models are four objects from  
\cite{claret} compilation. Their spectroscopic metallicity is consistent within errors with [Fe/H]=$-$0.40; 
the mass of the various components ranges between $\sim$1.4 and $\sim$4.2 $M_{\odot}$, and they are evolving along either 
the RGB or core He-burning 
phase. Figure~\ref{figClaretEB} compares their MR diagrams with theoretical isochrones for [Fe/H]=$-$0.40, that are able to match 
simultaneously both components (within their mass and radius error bars) in all four systems for the labelled ages, with our choices of 
MS core overshooting efficiency and mixing length. 

\begin{figure}[b]
% \vspace*{-2.0 cm}
\begin{center}
 \includegraphics[width=3.2in]{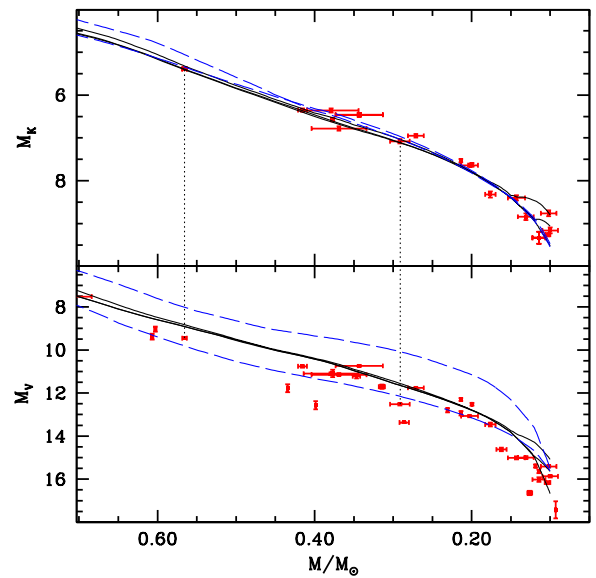} 
% \vspace*{-1.0 cm}
 \caption{Comparison of theoretical and observed mass-luminosity relationships in the $V$ and $K$ bands for a 
sample of low-mass stars from \cite{delfosse}. Filled symbols with error bars display the data, whilst 
solid lines correspond to three isochrones with [Fe/H]=0.06 and ages equal to 300~Myr, 
1~Gyr and 10~Gyr respectively. The dashed lines show 10~Gyr isochrones with [Fe/H]=$-$0.40 (brighter 
at fixed mass compared to the [Fe/H]=0.06 isochrones) and [Fe/H]=0.45, respectively. 
Dotted lines highlight two objects that are inconsistent with the models in the $V$-band, 
but fully consistent in the $K$-band (see text for details).}
   \label{figdelfosse}
\end{center}
\end{figure}

Finally, we compare the mass-luminosity relationship predicted by our low-mass models in the $V$ and $K$ bands, 
with the data presented by \cite{delfosse}, based mainly on visual and interferometric pairs. We display in 
Fig.~\ref{figdelfosse} the observational data together with three isochrones with [Fe/H]=0.06 and ages equal to 300~Myr, 
1~Gyr and 10~Gyr respectively (solid lines), plus two 10~Gyr isochrones with [Fe/H]=$-$0.40 and [Fe/H]=0.45 
respectively (dashed lines).
 
First of all, as also noted by \cite{delfosse}, the $V$-band data show a large dispersion at fixed $M$, with models 
matching a sort of upper envelope of the data. 
The $K$-band data are much tighter, and in very good general agreement with the [Fe/H]=0.06 models, even though the sample is smaller than for the $V$-band. 
It is interesting to consider the two objects highlighted by the dotted lines. They have estimates of both $V$ and $K$ 
absolute magnitudes; in the $K$-band the agreement with theory for [Fe/H]=0.06 (or higher) is essentially perfect, whereas in the $V$-band 
the data are clearly under-luminous compared to the models.
The fact that the $V$-band diagram is much more sensitive to the exact metallicity of the sample (as shown 
by the dashed lines in the figure) suggests that [Fe/H] may play a role in explaining this dispersion. The [Fe/H]=0.45 isochrone is 
under-luminous at fixed mass compared to the [Fe/H]=0.06 one, but still cannot explain the full dispersion of the data.

\begin{figure}[b]
% \vspace*{-2.0 cm}
\begin{center}
 \includegraphics[width=3.2in]{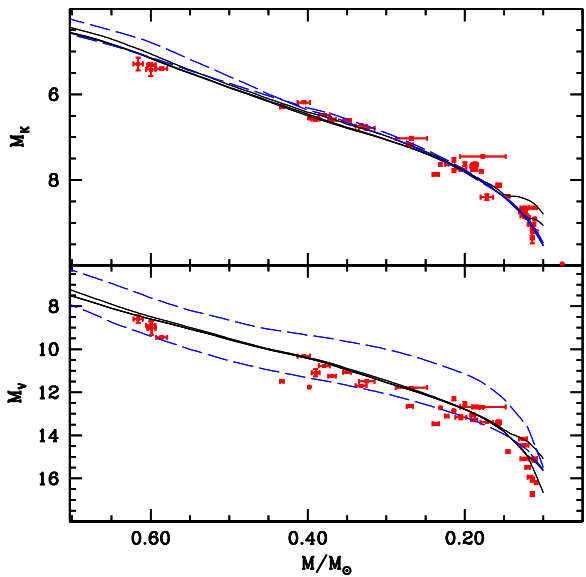} 
% \vspace*{-1.0 cm}
 \caption{As Fig.~\ref{figdelfosse} but for observed mass-luminosity relationships from \cite{benedict}  
(see text for details).}
   \label{figbenedict}
\end{center}
\end{figure}

{\bf Figure~\ref{figbenedict} displays a similar comparison with the more recent mass-luminosity empirical data by \cite{benedict}. 
Also in this case the dispersion in the $V$-band is larger than in the $K$-band. In the $K$-band (weakly sensitive to 
chemical composition) the agreement with theory is again generally quite good, apart from the cluster of 
objects with mass around 0.6$M_{\odot}$, that appear somewhat underluminous with respect to the models, irrespective of the 
adopted metallicity between [Fe/H]=$-$0.4 and 0.45.}

\subsection{Stars with asteroseismic mass determinations}

A recent study by \cite{tayar} has provided a sample of over 3000 RGB stars with $T_{eff}$, mass (determined from asteroseismic scaling relations), 
surface gravity, [Fe/H] and $\alpha$-enhancement ([$\alpha$/Fe]) determinations from the updated APOGEE-{\sl Kepler} catalog. 
These stars cover a log(g) range between $\sim$3.3 and 1.1 (in cgs units), and $T_{eff}$ between $\sim$5200 and 3900~K, with 
the bulk of the stars having [Fe/H] between $\sim -$0.7 and $\sim$ +0.4~dex, and a maximum $\alpha$-enhancement 
typically around 0.25~dex. This sample allows to compare empirically determined $T_{eff}$ values \citep[calibrated on the][temperature scale]{ghb09} 
with theoretical models of the appropriate chemical composition, that are very sensitive to the treatment of the superadiabatic layers, hence 
the calibration of $\alpha_{\rm ML}$.

\begin{figure}
   \centering
   \includegraphics[width=3.2in]{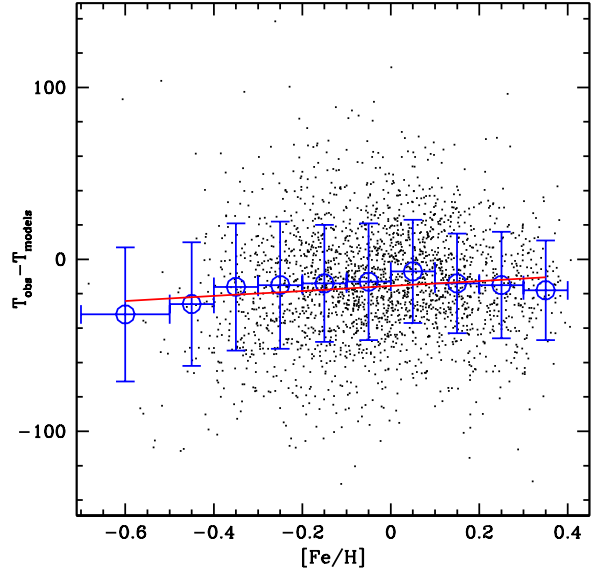}
      \caption{$\Delta T$ as a function of [Fe/H] (dots) for RGB stars with asteroseismic mass determinations from \cite{tayar}, and  
      ${\rm [\alpha/Fe]<0.07}$. Open circles with error bars denote the mean values of $\Delta T$  
      in specific metallicity bins, while the solid line displays a linear fit to the binned data. Vertical error bars denote 
      the 1$\sigma$ dispersion of around the mean values of $\Delta T$ in each bin, whereas the horizontal error bars denote the width of the 
      [Fe/H] bins.}
         \label{dttayar}
   \end{figure}

In our comparison we have considered only stars with [$\alpha$/Fe]$<$0.07 (this upper limit corresponds to  
approximately 3-5 times the quoted 1$\sigma$ error on [$\alpha$/Fe]), but an upper limit closer to zero 
does not change our results.  
We have calculated differences $\Delta T\equiv T_{obs}-T_{models}$ between observed and theoretical $T_{eff}$ 
for each individual star, by interpolating linearly in mass, [Fe/H] and log(g) amongst our models, 
to determine the corresponding theoretical $T_{eff}$.
{\bf The $\Delta T$ values for [Fe/H] larger than $\sim -$0.7~dex have been collected   
in ten [Fe/H] bins with total width of 0.10~dex, apart from the most metal poor one, that has a 
width of 0.20~dex, due to the smaller number of stars populating that metallicity range. We have then  
performed a linear fit to the mean $\Delta T$ values of each bin, and derived 
a slope equal to 14 $\pm$ 11 K/dex, statistically different from zero at much less than 2$\sigma$ (see Fig.~\ref{dttayar}). 
The average $\Delta T$ is equal to just $-$14~K, with a 1$\sigma$ dispersion of 34~K. This small offset between models and observations 
is well within the error on the \citet{ghb09} $T_{eff}$ calibration  
(the quoted average error on their RGB $T_{eff}$ scale is $\leq$76~K).}

%--------------------------------------------------->
%---------------------------------------------------|

\subsection{Star clusters}

The following comparisons with CMDs of a sample of Galactic open clusters and one globular cluster 
(with solar scaled initial metal distribution) provide additional tests of the reliability of our 
evolutionary tracks/isochrones plus the adopted bolometric corrections. 
In all these comparisons we have included the effect of extinction according to the standard \cite{ccm89} 
reddening law, with $R_V\equiv A_V/E(B-V)$=3.1.

\begin{figure}[b]
% \vspace*{-2.0 cm}
\begin{center}
 \includegraphics[width=3.2in]{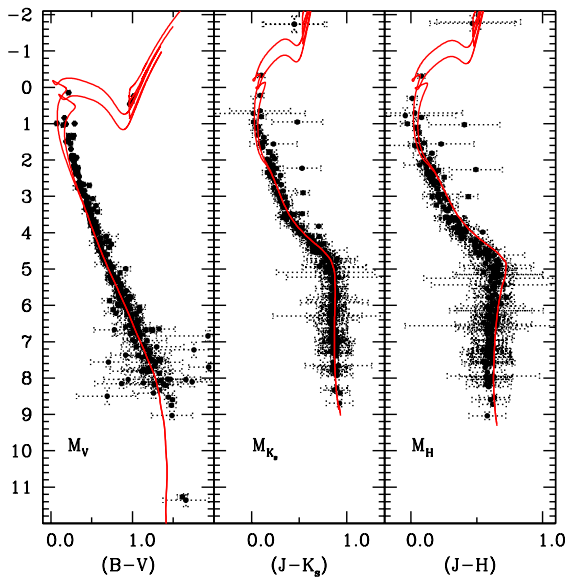} 
% \vspace*{-1.0 cm}
 \caption{Comparison between our 600 and 800~Myr, [Fe/H]=0.06 isochrones, and three Hyades CMDs, corrected 
for the  secular parallaxes determined by \cite{roser11} --see text for details.}
   \label{figHyadesbvjhk}
\end{center}
\end{figure}

Figure~\ref{figHyadesbvjhk} displays $BVJHK_s$ CMDs ($JHK_s$ from 2MASS photometry) for Hyades members 
taken from \cite{roser11} and \cite{kopytova16}, that reach the VLM star regime, down to 
$\sim$0.2$M_{\odot}$. We have calculated absolute magnitudes by applying the secular parallaxes 
determined by \cite{roser11}. The average parallax of these objects is in agreement with the 
average value of 103 probable members of the Hyades from the $Gaia$ data release 1, as given by \cite{gaia17}, 
within the quoted errors. We display also color and absolute magnitude error bars (the error bars on the 
absolute magnitudes account also for the contribution of the parallax errors) 
given that color errors often are non negligible along the MS.

The cluster CMDs are compared with our {\bf $t$=600 and 800~Myr}, [Fe/H]=0.06 isochrones 
-- close to spectroscopic estimates [Fe/H]=0.14$\pm$0.05 \citep{cayrel} and [Fe/H]=0.10$\pm$0.01 \citep{bj05} 
-- assuming $E(B-V)$=0, consistent with the results by \cite{taylor06}. {\bf The age range bracketed by these two isochrones 
is representative of the range of ages estimated for this cluster, as recently debated in the literature 
\citep[see, e.g.][and references therein]{perryman, brandt}].}

The theoretical isochrones follow well the observed MS down to the faintest limit, 
apart from the $JH$ diagram, that shows a systematic 
offset due to the $H$-band bolometric corrections, although models are still consistent 
with the data within the associated error bars.

\begin{figure}[b]
% \vspace*{-2.0 cm}
\begin{center}
 \includegraphics[width=3.1in]{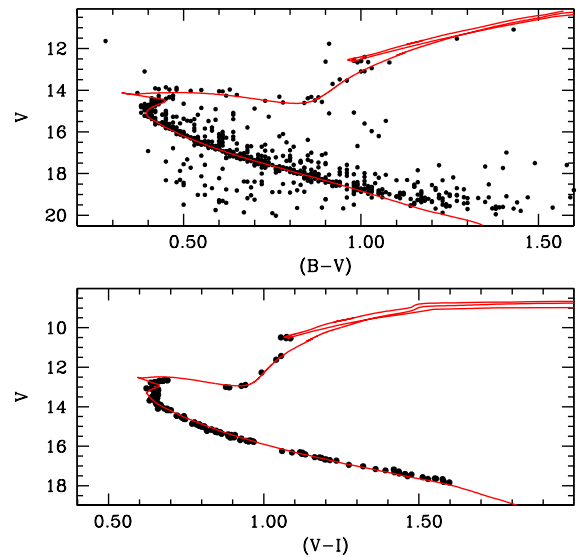} 
% \vspace*{-1.0 cm}
 \caption{Optical CMDs for NGC~2420 (top panel) and M~67 (bottom panel).  
Our isochrones with [Fe/H]$-$0.40, $t$=2.5~Gyr, $(m-M)_0$=11.95, $E(B-V)$=0.06 (top panel) and 
[Fe/H]=0.06, $t$=4~Gyr, $(m-M)_0$=9.64, $E(B-V)$=0.02 (bottom panel) 
are also shown (see text for details).}
   \label{figN2420M67}
\end{center}
\end{figure}

Optical CMDs of NGC~2420 \citep{twarog} and M~67 \citep{sandq} are shown in Fig.~\ref{figN2420M67}, compared to 
isochrones with $t$=2.5~Gyr and [Fe/H]=$-$0.40 in case of NGC~2420, $t$=4~Gyr and [Fe/H]=0.06 for M~67, respectively. 
These metallicities are consistent with [Fe/H]=$-$0.44$\pm$0.06 (NGC2420) and [Fe/H]=0.02$\pm$0.06 (M67) quoted by 
\cite{gratton00}.
The isochrones have been shifted to account for distance moduli 
and reddenings $(m-M)_0$=11.95, $E(B-V)$=0.06 for NGC~2420, and $(m-M)_0$=9.64, $E(B-V)$=0.02 for M~67. 
These pairs of values are consistent with the reddening estimates by \cite{twarog97} and MS-fitting 
distance moduli (using dwarfs with accurate Hipparcos parallaxes) by \cite{ps03}, within their error bars.

The values of the mass evolving at the TO for NGC~2420 and M~67 isochrones are $\sim$1.3$M_{\odot}$ 
and $\sim$1.2$M_{\odot}$ respectively, in the mass range where the size of the overshooting region 
is decreased down to zero from the standard value of 0.2$H_p$. The shape of the TO region --sensitive 
to the extent of the overshooting region-- is well traced by the isochrones for both clusters, 
lending some support to our prescription for the reduction of size of the overshooting region with mass.

One can notice also how, in addition to the MS (apart from the faintest end of NGC~2420 MS), 
also RGB, SGB and red clump sequences are nicely matched by the isochrones.

The next object compared to our isochrones is the old and super metal rich open cluster NGC~6791. 
At the super-solar metallicity of this object, bolometric corrections are bound to be more uncertain, 
because inaccuracies in atomic and molecular opacity data entering the model spectra calculations 
are greatly enhanced in this metallicity regime. 

For this cluster we take advantage of the analysis by \cite{brogaard11} and \cite{brogaard12} 
of two DEB systems, that 
provide estimates of $E(B-V)$=0.16$\pm$0.025, $(m-M)_V$=13.51$\pm$0.06, 
[Fe/H]=+0.29$\pm$0.03(random)$\pm$0.07(systematic), this latter value in agreement, within the errors,  
with spectroscopic estimates by \cite{ovrf06} and \cite{carraro06}, but lower than 
[Fe/H]=+0.47$\pm$0.04 determined by \cite{gratton06}. 

\begin{figure}[b]
% \vspace*{-2.0 cm}
\begin{center}
 \includegraphics[width=3.1in]{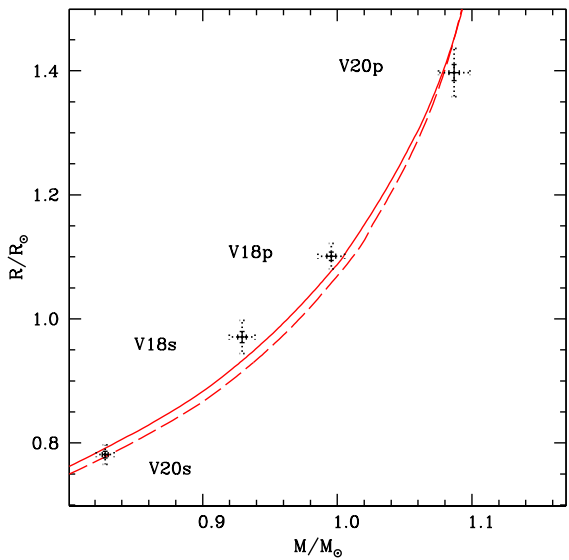} 
% \vspace*{-1.0 cm}
 \caption{Comparison in the MR diagram between the primary and secondary components of the 
DEB systems V18 and V20 in NGC6791, and our  
[Fe/H]=0.30 isochrones with $t$=8.5~Gyr including atomic diffusion (solid line), and $t$=9.0~Gyr 
without atomic diffusion (dashed line).}
    \label{figN6791EB}
\end{center}
\end{figure}

Figure~\ref{figN6791EB} displays the MR diagram for the four components (the primary component of 
V20 is in the TO region of the CMD, the other components are increasingly fainter MS stars) of these 
two DEB systems 
\citep[named V18 and V20 in][]{brogaard11} including the 1$\sigma$ and 
3$\sigma$ error bars, together with two theoretical isochrones for [Fe/H]=0.30, with and without the inclusion of atomic diffusion\footnote{
When diffusion is efficient, the quoted isochrone [Fe/H] corresponds to the initial value, that 
is also the one reinstated along the RGB by the deepening convection, after the first dredge up is 
completed. Notice that the spectroscopic measurements of [Fe/H] in NGC~6791 and the Galactic globular cluster Rup~106 discussed later,  
have been obtained for bright RGB stars.}, and ages of 8.5~Gyr and 9.0~Gyr, respectively.

As for the isochrones discussed by \cite{brogaard12} it is not possible to match perfectly 
the MR diagram of these EBs with theoretical isochrones. Those shown in Fig.~\ref{figN6791EB} represent 
the best compromise to match the data for the four DEB components, within their errors. 

\begin{figure}[b]
% \vspace*{-2.0 cm}
\begin{center}
 \includegraphics[width=3.1in]{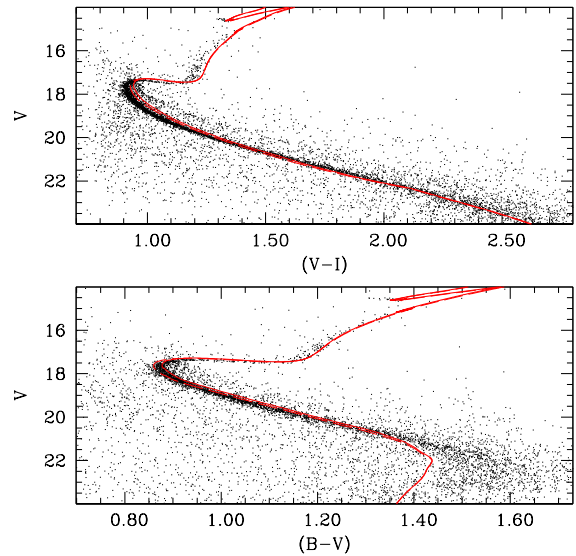} 
% \vspace*{-1.0 cm}
 \caption{$BVI$ CMDs for NGC~6791, compared to the same isochrones of Fig.~\ref{figN6791EB}.  
The isochrones have been shifted in magnitude and colors by $(m-M)_V$=13.52 (isochrone with diffusion, shown as 
a solid line) and $(m-M)_V$=13.54 (isochrone without diffusion, shown as a dashed line), and $E(B-V)$=0.16 
(see text for details).}
   \label{figN6791CMD}
\end{center}
\end{figure}

\begin{figure}[b]
% \vspace*{-2.0 cm}
\begin{center}
 \includegraphics[width=3.0in]{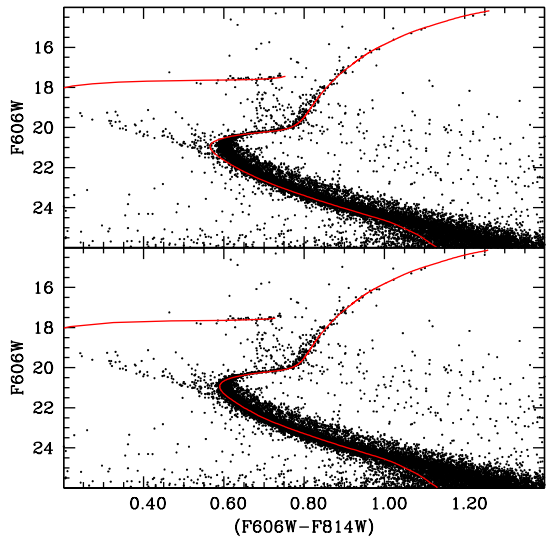} 
% \vspace*{-1.0 cm}
 \caption{Optical $HST$/ACS CMD for Rup~106, compared to our ZAHB sequences and isochrones with 
[Fe/H]=$-$1.55, $t$=12.5~Gyr, $(m-M)_0$=16.69, $E(B-V)$=0.18 and no atomic diffusion (top panel), 
and $t$=11.5~Gyr $(m-M)_0$=16.66, $E(B-V)$=0.18, including atomic diffusion (lower panel -- see text for details).}
   \label{figRup106}
\end{center}
\end{figure}

Figure~\ref{figN6791CMD} places the same isochrones of the DEBs comparison in optical $BVI$ CMDs, together 
witj the cluster photometry, corrected for differential reddening, by \cite{brogaard12}. 
We have displayed only stars with good quality photometry, i.e. we 
have considered only objects with photometric reduction yielding a 
$sharp$ index between -0.4 and +0.4, and a $chi$ index between 0.9 and 1.2. 
The isochrones have been shifted in colour for a reddening $E(B-V)$=0.16, and 
vertically for $(m-M)_V$=13.52 (the isochrone with diffusion) and $(m-M)_V$=13.54 
(the isochrone without diffusion) respectively.
These distance moduli, both consistent with the result from the DEB analyses, 
allow to match the $V$-band magnitude of the 
observed red clump stars with the core-He-burning portion of the isochrones. The overall 
comparison is better in the $BV$ CMD, where the RGB location and slope is reasonably reproduced, as well as
the TO-SGB-upper MS sequence. The TO region is matched better by the isochrone including atomic diffusion.
In $VI$ the match is overall worse. The RGB of the isochrones is redder than observed, the TO-SGB region is less well 
reproduced than in $BV$, although the lower MS is better matched.

As a last object, we have considered the low-mass \citep[total actual mass lower than $10^5 M_{\odot}$, see][]{villanova13} 
outer halo Galactic globular cluster Rup~106, whose stars display a solar scaled metal distribution, without 
the O-Na and C-N abundance anticorrelations common in other Galactic globular clusters \citep{villanova13}. 
Figure~\ref{figRup106} shows the cluster optical CMD in the $HST$/ACS camera photometric system \citep{dsa11}, together 
with isochrones for [Fe/H]=$-$1.55 --close to the mean value [Fe/H]=$-$1.47$\pm$0.02 determined 
spectroscopically by \cite{villanova13} --  $t$=12.5~Gyr (without atomic diffusion) and $t$=11.5~Gyr (including atomic diffusion), and ZAHB sequences 
(obtained from models with and without diffusion, respectively) for the same 
metallicity. A reddening $E(B-V)$=0.18 and distance moduli $(m-M)_0$=16.66 (for isochrones and ZAHB models 
with diffusion) and $(m-M)_0$=16.69 (for isochrones and ZAHB models without diffusion) have been applied to 
the models. The distance moduli have been fixed by matching the theoretical ZAHB sequences to the lower envelope of 
the observed HB.

The isochrones follow well the observed CMD. The TO region is better matched by the isochrone including 
atomic diffusion. Increasing the age of the isochrone without diffusion to make its TO redder 
does not improve the match with 
observations, because the model SGB would become fainter than the observed one.

\section{Conclusions}
\label{conclusions}

In this paper we have presented a comprehensive  overview of the updated BaSTI models, discussing the change in physics inputs compared 
to the previous BaSTI calculations, including comparisons with recent independent stellar model and isochrone databases, and a host 
of observational tests.
Improving upon the previous BaSTI release, this new library increases significantly the number of available metallicities, includes also the VLM regime, 
accounts consistently for the pre-MS evolution in the isochrone calculations, and provides also asteroseismic properties of the models.

Our new models/isochrones are able to match several sets of independent observational constraints 
that involve pre-MS stars and objects in more advanced evolutionary phases, either single, in DEBs or in star clusters. We believe that this 
updated BaSTI release will be an important tool to investigate field and cluster, Galactic and extragalaxy stellar populations. 

We make publicly available the whole database of models and isochrones through a dedicated Web site at the following URL address: http://basti-iac.oa-abruzzo.inaf.it.
Here we provide tables of stellar evolutionary 
tracks and asteroseismic properties of our grid of stellar evolution calculations plus isochrones, in several photometric systems.
We can also provide, upon request, additional calculations and both evolutionary and asteroseismic outputs, for stellar masses
not in our standard grids.

In the near future we will set up a Web interface to enable interpolations  
in metallicity within the available track and isochrone grids, as well as 
the calculations of isochrones and  luminosity functions for any specified age.

The next paper of this series will present $\alpha-$enhanced and $\alpha-$depleted models and isochrones, particularly  
suited to study stellar populations in globular clusters and dwarf galaxies.

\acknowledgements{SC acknowledges partial financial support from PRIN-INAF2014 (PI: S. Cassisi) and from \lq{Progetto Premiale}\rq\ MIUR {\sl MITIC} (PI: B. Garilli).
Funding for the Stellar Astrophysics Centre is provided by The Danish National Research Foundation (Grant agreement no.: DNRF106). 
V.S.A. acknowledges support from the Villum Foundation (Research grant 10118). This research has been supported by the Spanish Ministry of Economy and
Competitiveness (MINECO) under the grant SEV-2011-0187 (A. Aparicio \& S. Hidalgo). We warmly thank F. Castelli for her helpful comments and suggestions.
We warmly thank the referee for a prompt and extremely helpful report, that has greatly improved 
the presentation of our results.}

\bibliographystyle{aasjournal.bst}
\bibliography{basti_new}

\end{document}